\documentclass[aps,pre,superscriptaddress]{revtex4-1}
\usepackage{epsfig,graphics,amssymb,amsmath,subeqnarray,mathrsfs,color,xcolor}
\usepackage[colorlinks=true, citecolor=red, linkcolor=blue,urlcolor=blue ]{hyperref}

\def\r{\mathbf{r}}
\def\br{\bar{\mathbf{r}}}
\def\ui{^{(i)}}
\def\uj{^{(j)}}
\def\um{^{(n)}}
\def\ua{^{(1)}}
\def\ub{^{(2)}}
\def\f{\mathbf{f}}
\def\t{\mathbf{t}}
\def\bt{\bar{t}}
\def\I{\mathbf{I}}
\def\v{\mathbf{v}}
\def\e{\mathbf{e}}
\def\d{\mathbf{d}}
\def\bv{\bar{\mathbf{v}}}
\def\uji{^{(j)\to(i)}}
\def\bs{\bar{s}}
\def\R{\mathbf{R}}

\def\hx{\hat{x}}
\def\hy{\hat{y}}
\def\z{\mathbf{z}}

\def\Bu{{\rm Bu}}
\renewcommand{\eqref}{Eq.~\originaleqref}

\begin{document}
\title{Bundling of elastic filaments induced by hydrodynamic interactions}
\author{Yi Man\footnote{Current address: Department of Aerospace and Mechanical Engineering, University of Southern California, USA.}}
\affiliation{
Department of Applied Mathematics and Theoretical Physics, 
University of Cambridge, CB3 0WA, United Kingdom.}
\author{William Page}
\author{Robert J. Poole}
\affiliation{School of Engineering, University of Liverpool, Liverpool, L69 3GH, United Kingdom.}
\author{Eric Lauga}
\email{e.lauga@damtp.cam.ac.uk}
\affiliation{
Department of Applied Mathematics and Theoretical Physics, 
University of Cambridge, CB3 0WA, United Kingdom.}
\date{\today}
\begin{abstract}

Peritrichous bacteria swim in viscous fluids by rotating multiple helical flagellar filaments. As the bacterium swims forward, all its flagella rotate in synchrony behind the cell in a tight helical bundle. When the bacterium changes its direction, the flagellar filaments unbundle and randomly reorient the cell for a short period of time before returning to their bundled state and resuming swimming. This rapid bundling and unbundling is, at its heart, a mechanical process whereby hydrodynamic interactions balance with elasticity to determine the time-varying deformation of the filaments. Inspired by this biophysical problem, we present in this paper what is perhaps the simplest model of bundling whereby two, or more, straight elastic filaments immersed in a viscous fluid rotate about their centreline, inducing rotational flows which tend to bend the filaments around each other. We derive an integro-differential equation governing the shape of the filaments resulting from mechanical balance in a viscous fluid at low Reynolds number. We show that such equation may be evaluated asymptotically analytically in the long-wavelength limit, leading to a local partial differential equation governed by a single dimensionless Bundling number. A numerical study of the dynamics predicted by the model reveals the presence of two configuration instabilities with increasing Bundling numbers: first to a crossing state where filaments touch at one point and then to a bundled state where filaments wrap along each other in a helical fashion. 
We also consider the case of multiple filaments, and the unbundling dynamics. We next provide an intuitive physical model for the crossing instability and show that it may be used to predict analytically its threshold, and adapted to address the transition to a bundling state. 
We then use a macro-scale experimental implementation of the two-filament configuration in order to validate our theoretical predictions and obtain excellent agreement. 
This long-wavelength model of bundling will be applicable to other problems in biological physics and provides the groundwork for further, more realistic, models of flagellar bundling.

\end{abstract}
\maketitle
\section{Introduction}

The locomotion of bacteria has recently provided the fluid mechanics community with a series  of outstanding problems at the intersection of many fields \cite{lauga16}. Building on classical work from the 1970's quantifying how bacteria actuate their helical flagella and how the forces from the fluid affect the kinematics of the organism \cite{chwang71,lighthill75}, recent work has addressed how to improve these classical models \cite{lighthill96_theorem,lighthill96_helical}, how to predict and measure the flow induced by bacteria \cite{Watari10,spagnolie2011,drescher11}, and the crucial role that  hydrodynamics has played in the evolution of the bacterial flagella \cite{spagnolie2011}.  One aspect in particular which has received a lot of attention is the role of hydrodynamic interactions, including interactions with surfaces \cite{lauga06_circles,berke08,lemelle_2010,giacche10,dileonardo_2011,drescher11,spagnolie12,lopez14}, external flows \cite{sokolov09,marcos12,rusconi14}, complex fluids \cite{zhou14} and between cells  \cite{ishikawa07_bacteria,koch,sokolov12}.

\begin{figure}
\centering
\includegraphics[width=0.7\textwidth]{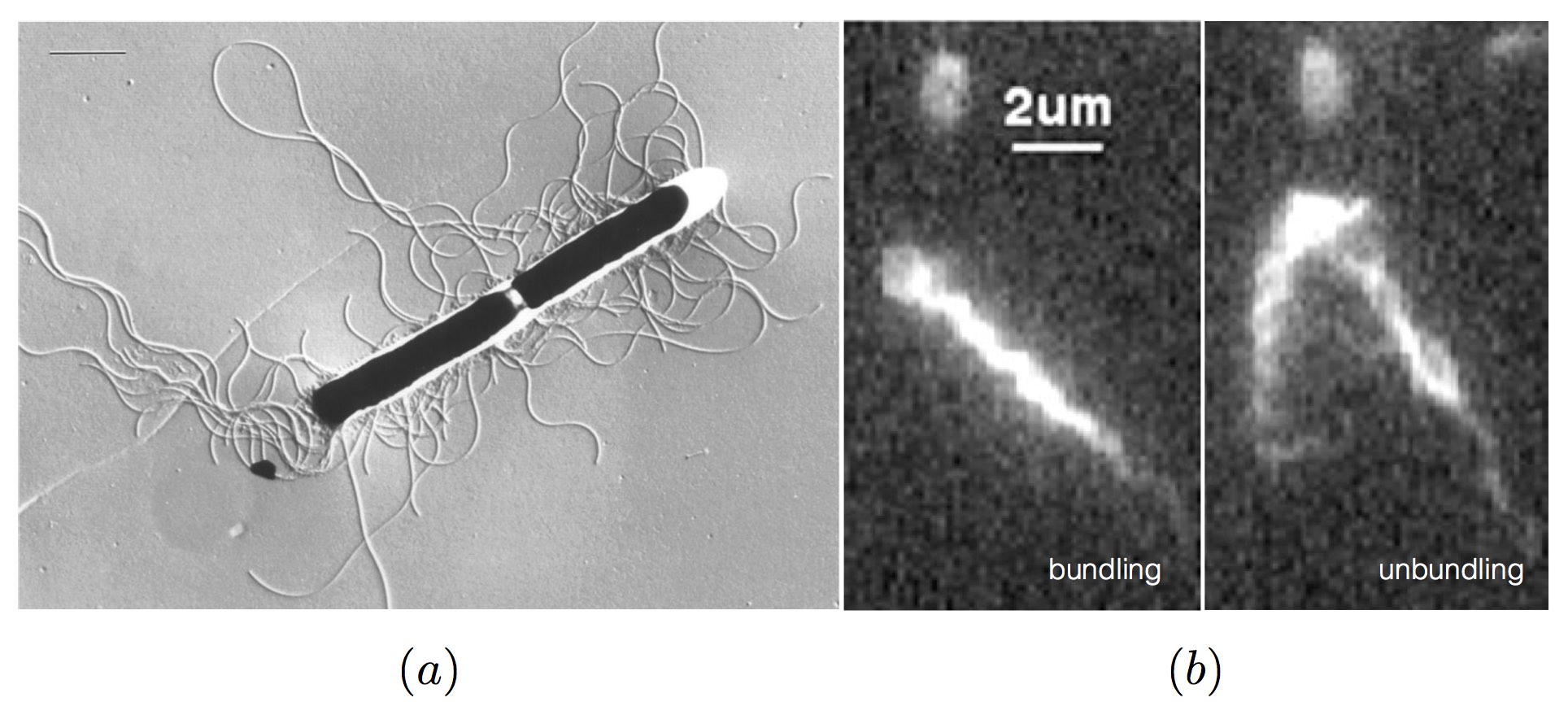}
\caption{Illustration of peritrichously flagellated bacteria; 
($a$)  {\it Bacillus subtilis} cell dividing \cite{Cisneros07};  ($b$)  Bundling and unbundling dynamics of bacterial flagellar filaments of {\it E.~coli} \cite{Turner00}. Panel (a) courtesy of
C. Dombrowski and R.E. Goldstein; Panel (b) reproduced from Turner et al. (2000) 
Real-time imaging of fluorescent flagellar filaments, {\it J.~Bacteriol},  {\bf 10}, 2793-2801, Copyright 2000 Society of Microbiology.}
\label{images}
\end{figure}

One particular phenomenon involving hydrodynamic interactions at the scale of a single cell is the bundling and unbundling of bacterial flagella \cite{Macnab77}.  While many bacteria have only one flagellum, most well-studied pathogenic bacteria possess multiple flagella, and  are refereed to as  ``peritrichous'' bacteria.  Such  bacteria are propelled from behind by a bundle of helical flagella, for example  the well-studied {\it Escherichia coli} ({\it E.~coli}) \cite{Berg73}, and also {\it Salmonella typhimurium} \cite{Macnab77}, {\it Halobacterium Salinarium} \cite{Kupper94}   or {\it Bacillus subtilis}  \cite{Cisneros07}  (see Fig.~\ref{images}a).

The presence of multiple flagella allow bacteria to undergo  random walk-like trajectories where long straight swimming `runs'  ($\sim$~1~s) are interrupted by short `tumbles' ($\sim$~0.1~s) during which  the cells  randomly reorient \cite{Berg73,Macnab77,Macnab83}. During the run phase, all flagella take a normal left-handed shape and are rotated in a counter-clockwise direction  (CCW, when viewed from outside the cell behind the flagella) by specialised rotatory motors embedded in the cell wall. During a tumble, at least one motor switches its rotation direction to clockwise (CW), the corresponding flagellar filaments fly out of the bundle  (`unbundling') and change their helical pitch, amplitude and handedness in a manner which is dictated by biochemistry \cite{Calladin75,Hasegawa98,Turner00}. This in turn leads  to a modification of the force balance on the whole cell and its reorientation. At the end of a tumble, all motors return to a CCW rotation, and the splayed flagella reintegrate into the normal left-handed helical bundle behind the cell (`bundling'), which resumes swimming along a straight line. 
This  bundling and unbundling of   flagellar filaments is illustrated in Fig.~\ref{images}b for {\it E.~coli}.

The process by which the flagella of peritrichous bacteria interact, repeatedly come together and separate is complex and involves at least three important mechanical aspects: 
     hydrodynamic interactions between the rotating helical filaments \cite{Macnab77, Kim03, Flores05, Reichert05, Reigh12, Lim12, Reigh13, Kanehl14, Janssen11}, elastic  deformation of the short  flexible hook joining the rotary motors to the flagellar  filaments  \cite{Reichert05,Brown12}
and  interactions between the filaments and   the rotating body of the cell \cite{Powers02, Adhyapak15}.

Given its relevance to one of the most fundamental forms of mobility on the planet, the role of fluid dynamics in this process has received a lot of attention from the research community. Experimentally, a macro-scale version of the interactions between rotated helices \cite{Kim03}, and  subsequent flow measurements \cite{Kim04},  showed that hydrodynamic interactions alone were able to lead to wrapping of flexible helical filaments consistent with experimental observations at the cellular scale \cite{Turner00}. Computationally, the issue of synchronisation between rotating helices was addressed  \cite{Kim04,Reichert05}, and similarly for rotating paddles \cite{Qian09}. Full simulations of the bundling process were carried using a variety of computational methods including the use of regularized flow singularities \cite{Flores05}, 
multi-particle collision dynamics \cite{Reigh12, Reigh13}, finite differences \cite{Janssen11}, the immersed boundary method  \cite{Lim12}, beads-spring models \cite{Watari10} and the boundary element method \cite{Kanehl14}.

In parallel to these significant computational advances, theoretical studies have not yet been able  to derive simplified models allowing one to capture, from first principle,  the essence of the dynamics of the bundling and unbundling process. In this paper, we derive the tools required to build such a model by considering what is perhaps the simplest model of bundling, namely the rotation and interaction of two rotated straight filaments.   This is clearly an idealised configuration  which ignores the helical geometry of real flagella and thus the propulsive flow generated by the rotation of the  filaments, but it has the advantage that it  may be treated  rigorously from a mathematical standpoint and provides the basis to develop more realistic models. 
Taking advantage of a separation of length scales for slender filaments we show that in the long-wavelength limit the shapes  of the filament, obeying a balance between hydrodynamic and elastic forces, satisfy a local nonlinear partial differential equation. Crucial to the derivation of this equation is the integration of non-local hydrodynamic interactions which can be done analytically  in the long-wavelength limit \cite{goldstein16,man2016hydrodynamic}. We then study the dynamics predicted by our model numerically and reveal the presence of two configuration instabilities, first to a crossing state where filaments touch at one point and then to a bundled state where filaments wrap along each other in a helical fashion.   Using a simplified analytical approach, we  are next able to rationalise and predict the onset of the instabilities.  We then use a macro-scale experimental implementation of the two-filament configuration in order to validate our theoretical  predictions and we obtain excellent agreement. Our derivations, which provide a simplified approach to capture the dynamics of bundling and unbundling, should be applicable to a wide range of problems in biological mechanics where slender filaments interact hydrodynamically.

\section{Interactions between two elastic filaments}
\subsection{Setup}\label{sec:setup}
\begin{figure}
\centering
\includegraphics[width=0.4\textwidth]{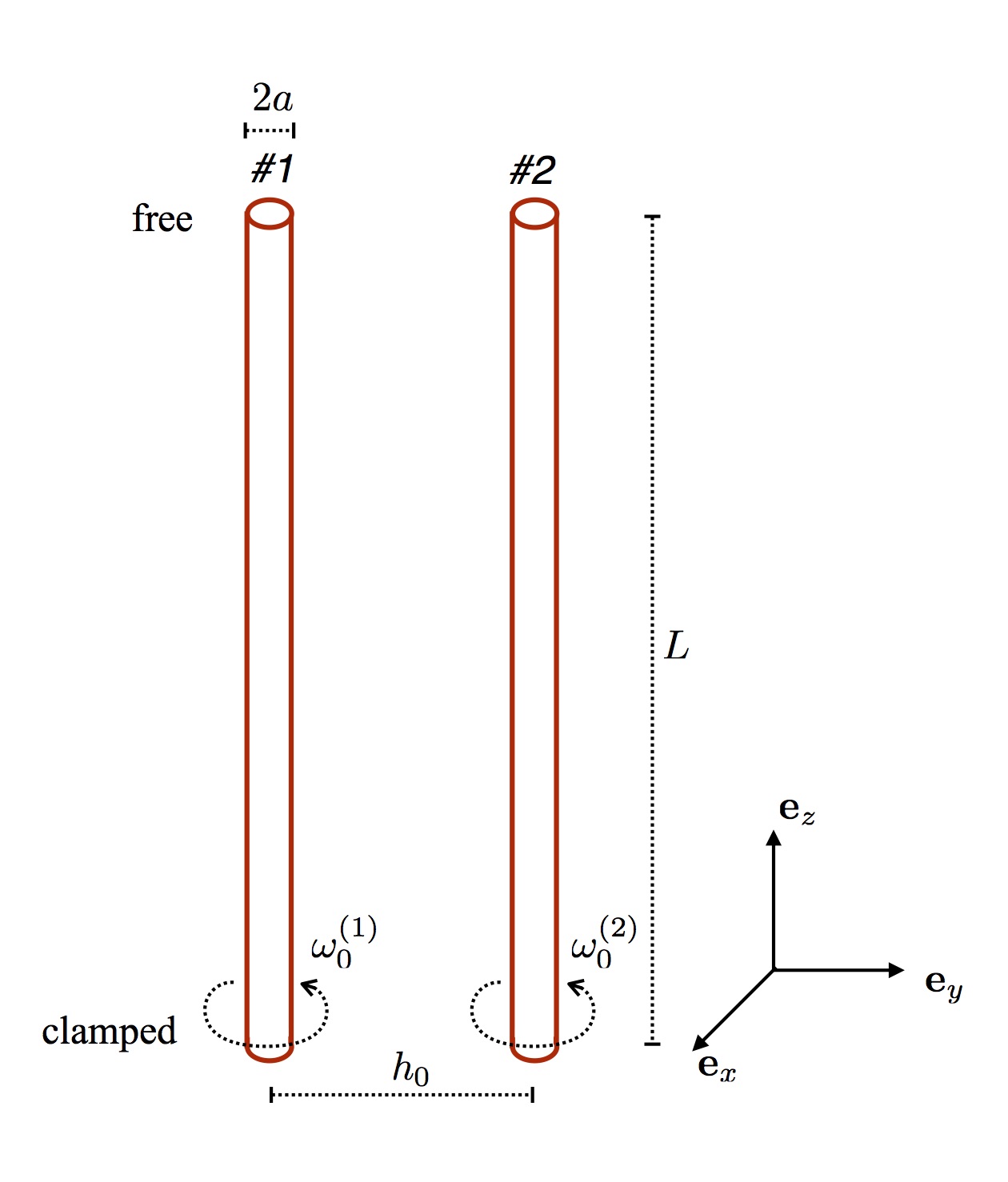}
\caption{A simplified model of flagellar bundling. Two parallel elastic filaments of length $L$ and radius $a$ are separated by a mean distance $h_0$. Both filaments are rotated with speed  $\omega_0\ui$ ($i=1,2$) in a viscous fluid from their clamped end while their other end remains mechanically free. We use Cartesian coordinates with $z$ along the filaments and $x,y$ perpendicular to it.}
\label{geo}
\end{figure}

To address flagellar bundling as a balance between hydrodynamics and elasticity, we set up the very simplified model illustrated in Fig.~\ref{geo}. Two naturally straight elastic filaments of length $L$ and radius $a$ are rotated at their clamped ends by imposed rotation rates $\omega\ui_0$ where $i=1,2$ refers to the filament number. The filaments are initially parallel and separated by a distance $h_0$. The filaments are assumed to be located in a viscous fluid of viscosity $\mu$. As a result, their rotation  set up rotational  flows and, with the other ends of both filaments free, will lead to a helical wrapping driven purely by hydrodynamic interactions. Note that in general we allow the two rotation rates of the filaments to  be different; however, in most of the  cases studied below we will consider identical rotation rate, i.e.~$\omega\ui_0=\omega_0$. Note also that while we focus our initial derivation to two filaments, the results will generalised later in the paper to the case of $N\geq 2$ filaments.

In order to derive an analytical model of the dynamics, we need to make a further assumption. Examining  the typical dimensions of flagellar filament of bacteria, we see that the three relevant length scales $a$, $h_0$, and $L$ are often in the  limit where there is a clear scale separation, $a\ll h_0\ll L$. As an illustrative example, consider the flagellar filaments of {\it E.~coli}. Each filament has a typical length $L\approx 10~\mu$m  and radius  $a\approx10$~nm.  The cell body takes approximately the shape of a prolate  ellipsoid  of $1~\mu$m  width and $2~\mu$m  length \cite{Turner00, berg03}. Assuming  the same scale as the body size, the distance between the proximal ends of the flagellar filaments can be thus estimated to be around $h_0\approx 1~\mu$m.  These numbers lead therefore to   ratios $a/h_0\sim 10^{-2}$ and $h_0/L\sim 10^{-1}$, which are consistent with this  separation of scales.

Mathematically, placing ourselves in the limit $a\ll L$ means that each filament is slender and we will be able to make use of resistive-force theory to compute hydrodynamic forces. The limit $h_0\ll L$ means that the filaments are not in the far-field limit but in the opposite limit where all long-range hydrodynamic interactions have to be included. Furthermore, given that their separation will remain at most $h_0$, this means that we will be able to treat the problem in the long-wavelength limit, a crucial step to derive an analytical model. 
Finally, the limit $a\ll h_0$ means that lubrication stresses can be ignored and the flow can be accurately captured by a superposition of hydrodynamic singularities. 

\subsection{Force balance}

Focusing on small-scale systems (a few microns, as relevant to the bacterial world), we remain safely in the low-Reynolds number regime, and therefore the  force balance on each filament is written instantaneously as
\begin{equation}\label{FB}
	\f\ui_h(s, t)+\f\ui_e(s, t)=0,
\end{equation}
where $\f\ui_h(s, t)$ and $\f\ui_e(s,t)$ refer, respectively, to the  hydrodynamic and elastic forces per unit length acting along filament $i$ and where $s$ refers to the arclength along the filament ranging from $s=0$ (clamped end) to $s=L$ (free end). We denote the position vector of  filament $i$ as $\r\ui(s,t)$.

\subsection{Kinematics}

Assuming that all displacements remain on the  order of $h_0$, it is natural to  describe the shapes of the filaments using  Cartesian coordinates (Fig.~\ref{geo}) as
\begin{equation}\label{shape}
\r\ui(	s, t)=[h_0x\ui(s, t), h_0 y\ui(s, t), s]=\left[h_0\zeta\ui(s,t), h_0\left(\mp\frac{1}{2}+\eta\ui(s,t)\right),s\right],
\end{equation}
where  $[x\ui, y\ui]$ and $[\zeta\ui, \eta\ui]$ denote the filament position and displacement scaled by $h_0$, respectively.


\section{Calculation of elastic force density}

\subsection{Classical rod theory}
We first compute  the elastic force density, $\f_e$, arising in Eq.~\ref{FB}. Notation-wise, we drop for simplicity the  upper index $(i)$ and consider one specific filament.  For an inextensible elastic filament able to both bend and twist, it is a classical result that $\f_e$ contains three terms, namely  \cite{Wolgemuth00, powersReview10} 
\begin{equation}\label{hyperdiff}
\f_e = -A\frac{\partial^4\r}{\partial s^4}+C\left[\Omega\left(\frac{\partial\r}{\partial s}\times\frac{\partial^2\r}{\partial s^2}\right)\right]_{s}-\left[\Lambda\frac{\partial\r}{\partial s}\right]_{s}.
\end{equation}
In Eq.~\ref{hyperdiff} the coefficients $A, C$ are the bending and twist moduli of the filament, respectively,  $\Omega$ is the twist density and  $\Lambda$ is the Lagrange multiplier (tension) enforcing the  inextensibility of the filament. The conservation law for twist density (often referred to as the compatibility equation) is  \cite{powersReview10}
\begin{equation}\label{4}
\frac{\partial\Omega}{\partial t}=\frac{\partial\omega}{\partial s}+
\left(
\frac{\partial \r}{\partial s}\times\frac{\partial^2\r}{\partial s^2}
\right)
\cdot\left[\frac{\partial\r}{\partial t}\right]_{s},
\end{equation}
where $\omega$, a notation shorthand for $ \omega\ui(s,t)$, is the local rotation rate of the filament around its  centreline. In addition, the  local torque balance along the filament is written as
\begin{equation}\label{5}
C\frac{\partial\Omega}{\partial s}=\xi_r\omega,
\end{equation}
where 
$ \xi_r\approx4\pi\mu a^2$ is the local  drag coefficient for rotation around the filament centreline. Substituting Eq.~\ref{5} into Eq.~\ref{4} classically leads to a forced diffusion equation for the twist density as 
\begin{equation}\label{twist rate}
\frac{\partial\Omega}{\partial t}=\frac{C}{\xi_r}\frac{\partial^2\Omega}{\partial s^2}+
\left(\frac{\partial \r}{\partial s}\times\frac{\partial^2\r}{\partial s^2}
\right)
\cdot\left[\frac{\partial\r}{\partial t}\right]_{s}.
\end{equation}

\subsection{Separation of time scales}
In order to appreciate  the relative magnitude of each term in Eq.~\ref{hyperdiff}, we consider the physical scalings of the various quantities of interest. The relevant time scale in Eq.~\ref{twist rate} is a diffusion time, $T_t$, scaling as \cite{Wolgemuth00}
\begin{equation}
T_t=\frac{\xi_rL^2}{C}\cdot
\end{equation}
In contrast, balancing Eq.~\ref{hyperdiff} with an usual hydrodynamic drag term of the form $\sim \xi_\perp\partial \r/ \partial t$, reveals that the relevant bending time scale in Eq.~\ref{hyperdiff} is a hyper--diffusion time, $T_b$, scaling as \cite{WigginsGoldstein} 
\begin{equation}
	T_b=\frac{\xi_\perp L^4}{A}.
	\end{equation}
In the slender limit $L/a\gg 1$ we have the classical result for the drag coefficient \cite{cox70,Lauga09}
\begin{equation}
\xi_{\perp}\approx\displaystyle\frac{4\pi\mu}{\ln\left({L}/{a}\right)},
\end{equation}
and therefore, after introducing the ratio $\Gamma$ of elastic modulus,
\begin{equation}
\Gamma=\frac{C}{A},
\end{equation}
we obtain that the ratio of twist to bending relaxation time scales is given by the scaling
\begin{equation}\label{timescales}
\frac{T_t}{T_b}\sim \frac{\epsilon_a^2\ln(1/\epsilon_a)}{\Gamma},
\end{equation}
where we have defined $\epsilon_a\equiv a/L$, a small number in the slender limit. 
Clearly Eq.~\ref{timescales} indicates that $T_t \ll T_b$. Since $T_b$ is the time scale relevant to describe the shape of the filaments, we will be able to separate the time scales and consider the twist problem  solved in a  quasi-steady fashion \cite{Wolgemuth00}.  

\subsection{Non-dimensionalisation}

We proceed by nondimensionalising the equations using the length scale $L$ and time scale $T_b$ as the relevant dimensions and we use bars to denote dimensionless variables. 
Following Eq.~\ref{shape}, we write for the  position of the filament
 \begin{equation}\label{position}
\br=[\epsilon_h \bar x, \epsilon_h \bar y, \bs],
\end{equation}
where we have defined a second dimensionless number 
$ \epsilon_h\equiv {h_0}/{L}$, which is assumed to be small. Writing as well
\begin{equation}\label{ND_scalings}
\bs=s/L,  \quad \bar{\Lambda}=\Lambda \frac{L^2}{A}, \quad \bar{\f}_e=\f_e\frac{L^3}{A}, \quad \bar{\Omega}=\Omega L,
\end{equation}
we obtain the  dimensionless equations as
\begin{subeqnarray}
\slabel{NDa}		{\bar{\mathbf{f}}}_e &=&-\frac{\partial^4\br\ui}{\partial \bs^4}+\Gamma\left[\bar{\Omega}\left(\frac{\partial\br}{\partial \bs}\times\frac{\partial^2\br}{\partial \bs^2}\right)\right]_{\bs}-\left[\bar{\Lambda}\frac{\partial\br}{\partial \bs}\right]_{\bs},\\
\slabel{NDb}		\frac{\partial\bar{\Omega}}{\partial \bt}&=&\frac{\Gamma}{\epsilon_a^2\ln(1/\epsilon_a)}\frac{\partial^2\bar{\Omega}}{\partial \bs^2}+\frac{\partial \br}{\partial \bs}\times\frac{\partial^2\br}{\partial \bs^2}\cdot\left[\frac{\partial\br}{\partial \bt}\right]_{\bs}\cdot
\end{subeqnarray}
Note finally that the   imposed rotation is nondimensionalised  as $ \bar{\omega}_0=\omega_0T_b$, and the result is related to the classical dimensionless Sperm number, ${\rm Sp}$,  quantifying a balance between elastic and viscous drag as \cite{Lauga09}
 \begin{equation}\label{Sp}
 \bar{\omega}_0=\frac{\xi_\perp\omega_0  L^4}{A}\equiv \rm Sp^4.	
  \end{equation}

\subsection{Twist equilibrium}
\label{sec:twist}
For convenience, we now drop the `bar' notation in what follows. Except where explicitly stated, the results below should thus be assumed to be dimensionless. 

Given the separation of time scales by Eq.~\ref{timescales}, we expect  the first term on the right-hand of Eq.~\ref{NDb} to dominate and thus with the  boundary condition at the clamped and free ends, 
\begin{equation}
\displaystyle\frac{\partial\Omega}{\partial s}(s=0)=\frac{\epsilon_a^2\ln(1/\epsilon_a)}{\Gamma}\omega_0, \quad\Omega (s=1)=0, 
\end{equation}
the twist density  is given in quasi-steady equilibrium by a simple linear function
\begin{equation}\label{twist_rate}
 \Omega=\frac{\epsilon_a^2\ln(1/\epsilon_a)}{\Gamma}\omega_0(s-1).
\end{equation}
Substituting this result into Eq.~\ref{NDa},  we obtain the explicit formula for the elastic force density  as
\begin{equation}\label{fe_intermediate}
	\f_e=-\frac{\partial^4\r}{\partial s^4}+\omega_0\epsilon_a^2\ln\epsilon_a\left[(1-s)\left(\frac{\partial\r}{\partial s}\times\frac{\partial^2\r}{\partial s^2}\right)\right]_{s}-\left[\Lambda\frac{\partial\r}{\partial s}\right]_{s}.
\end{equation}
\subsection{Scalings}
In order to make further progress, we next compare the expected magnitude of each term in Eq.~\ref{fe_intermediate}. The first two terms clearly scale as
\begin{subeqnarray}\label{eq:scalings}
\slabel{eq:scalingsa}	\frac{\partial^4\r}{\partial s^4}&\sim& \epsilon_h,\\
\slabel{eq:scalingsb}	\omega_0\epsilon_a^2\ln\epsilon_a\left[(1-s)\left(\frac{\partial\r}{\partial s}\times\frac{\partial^2\r}{\partial s^2}\right)\right]_{s}&\sim& \epsilon_h\epsilon_a^2\ln(1/\epsilon_a){\rm Sp}^4.
\end{subeqnarray}
In order to derive the scaling for the third term, we need to carefully examine the equation for the tension, $\Lambda$. 

\subsection{Tension}

In order to solve for the  Lagrange multiplier $\Lambda$ explicitly, we consider the original equation for the force density,  Eq.~\ref{hyperdiff}. The inextensibility condition is mathematically written $\r_s\cdot \r_s=1$ or $\r_{ts}\cdot\r_s=0$, where we use superscripts to denote partial derivatives. We next compute the $s$ derivative of the force density as
\begin{align}
\begin{split}
\f_s&=\left(\I-\frac{1}{2}\r_s\r_s\right)\cdot\left(\r_{ts}-\v_s\right)-\frac{1}{2}(\r_{ss}\r_s+\r_s\r_{ss})\cdot(\r_t-\v)\\
&=\r_{ts}-\left(\I-\frac{1}{2}\r_s\r_s\right)\cdot\v_s-\frac{1}{2}(\r_{ss}\r_s+\r_s\r_{ss})\cdot(\r_t-\v),
\end{split}
\end{align}
and now aim to  simplify all terms involving time derivatives in Eq.~\ref{hyperdiff}. Since $\r_s\cdot \r_s=1$, it is clear that a derivative of this equation leads to $\r_s\cdot\r_{ss}=0$. Evaluating next the dot product $\f_s\cdot\r_s$ we have 
\begin{align}
\begin{split}
\f_s\cdot\r_s&=-\v_s\cdot\r_s+\frac{1}{2}\v_s\cdot\r_s-\frac{1}{2}\r_{ss}\cdot(\r_t-\v)\\
&=-\frac{1}{2}\v_s\cdot\r_s-\frac{1}{2}\r_{ss}\cdot(\r_t-\v).
\end{split}
\end{align}
We then calculate $\f\cdot\r_{ss}$ as
\begin{align}
\f\cdot\r_{ss}=\r_{ss}\cdot(\r_t-\v),
\end{align}
and can now eliminate all terms involving time derivatives by combining these equations as
\begin{equation}\label{eq:new}
2\f_s\cdot\r_s+\f\cdot\r_{ss}=-\v_s\cdot\r_s.
\end{equation}
Substituting the equation for the elastic force, Eq.~\ref{hyperdiff},  into Eq.~\ref{eq:new}, we obtain
\begin{equation}
2(\Lambda\r_s)_{ss}\cdot\r_s+(\Lambda\r_s)_s\cdot\r_{ss}=-2\r_{5s}\cdot\r_s+2\Gamma[\Omega(\r_s\times\r_{ss})]_{ss}\cdot\r_s-\r_{4s}\cdot\r_{ss}+\Gamma[\Omega(\r_s\times\r_{ss})]_s\cdot\r_{ss}+\v_s\cdot\r_s.
\end{equation}
The terms on the left-hand side can be simplified by noting that
\begin{align}
2(\Lambda\r_s)_{ss}\cdot\r_s+(\Lambda\r_s)_s\cdot\r_{ss}=2\Lambda_{ss}+\Lambda\r_{3s}\cdot\r_{s},
\end{align}
while the terms on the right-hand side involving the twist density combine into
\begin{align}
2\Gamma[\Omega(\r_s\times\r_{ss})]_{ss}\cdot\r_s+\Gamma[\Omega(\r_s\times\r_{ss})]_s\cdot\r_{ss}
=-\Gamma\Omega\r_s\times\r_{3s}\cdot\r_{ss}.
\end{align}
The equation for $\Lambda$ takes therefore the final form
\begin{equation}\label{eq:newLambda}
2\Lambda_{ss}+\Lambda\r_{3s}\cdot\r_s=-2\r_{5s}\cdot\r_s-\r_{4s}\cdot\r_{ss}-\Gamma\Omega\r_s\times\r_{3s}\cdot\r_{ss}+\v_s\cdot\r_s.
\end{equation}

Since $s=O(1)$, we see that the first term on the left-hand side of Eq.~\ref{eq:newLambda} provides the leading-order scaling for the magnitude of $\Lambda$. This needs to be balanced with the leading-order term on the right-hand side of the equation, which includes three terms scaling respectively as $O(\epsilon_h^2)$, $O(\epsilon_h^2\epsilon_a^2\ln(1/\epsilon_a)\rm Sp^4)$ and $O(\epsilon_h\v)$.  Since $\v$ is expected to scale  with $\f$, the term $\v_s\cdot\r_s$ will contribute at higher order, and therefore  we obtain the scaling for the final term in Eq.~\ref{fe_intermediate} as
\begin{equation}\label{eq:scalingLambda}
	\left(\Lambda\frac{\partial\r}{\partial s}\right)_{s} \sim O(\max\{\epsilon_h^2,\epsilon_h^2\epsilon_a^2\ln(1/\epsilon_a)\rm Sp^4\}).	 
\end{equation}

\subsection{Orders of magnitude and final scalings}

In order to estimate  the relative magnitude of the bending, twisting and tension terms in the case relevant to the bundling of bacterial flagella, we need to  examine the numbers applicable in the biological world.  Beyond the length scales mentioned above, we may use  past measurements for bacterial flagellar  filaments \cite{Hoshikawa85} to obtain the estimates
\begin{equation}
\omega_0\approx 100~{\rm s}^{-1},\quad A\approx 10^{-24}-10^{-22}~{\rm Nm}^2.
\end{equation}
These numbers  imply that the range in Sperm numbers is ${\rm Sp} \approx 3 -10$. 
As seen  in Section \ref{sec:setup}, a typical value for $\epsilon_a$ is around $10^{-3}$, and thus we see that
\begin{equation}\label{spermNo}
\epsilon_a^2\ln\left(1/\epsilon_a\right){\rm Sp}^4\ll 1.
\end{equation}
This in turn means that the tension in  Eq.~\ref{spermNo} scales in fact as 
\begin{equation}\label{eq31}
	\left(\Lambda\frac{\partial\r}{\partial s}\right)_{s} \sim O(	\epsilon_h^2),
\end{equation}
and that the term in Eq.~\ref{eq:scalingsb} can be neglected when compared with the one in 
Eq.~\ref{eq:scalingsa}. Since $\epsilon_h\ll 1$, the tension term in Eq.~\ref{eq31} may also be neglected in comparison with the bending term in Eq.~\ref{eq:scalingsa}. In the dynamical regime relevant to the helical filaments of bacteria, we thus have the final result
 \begin{equation}\label{fe_nd}
	\f_e=-\frac{\partial^4\r}{\partial s^4},
\end{equation}
 and  the elastic force density is  dominated by the bending term.

\section{Calculation of hydrodynamic force density}

Having evaluated the elastic force density in the limit relevant to the bundling of bacterial flagella, we here consider the second term appearing in Eq.~\ref{FB}, namely the hydrodynamic force density. We derive its value in the long-wavelength limit;  an early version of this calculation  was presented in Ref.~\cite{goldstein16}  in the linearised limit and 
Ref.~\cite{man2016hydrodynamic} in three dimensions.

\subsection{Resistive-force theory}

Since the filament is slender ($a\ll L$), the hydrodynamic force density is provided  by  resistive-force theory, which states that the force is  proportional to the local velocity of the filament relative to the background flow \cite{cox70,Lauga09}, as
\begin{equation}\label{rft}
\f_h\ui= -\left[\xi_\parallel \t\ui\t\ui +\xi_\perp\left(\I-\t\ui\t\ui\right)\right]\cdot\left(\frac{\partial\r\ui}{\partial t}-\v\uji\right),
\end{equation}
where $\v\uji$ denotes the flow velocity induced by the motion of filament $j$ near filament $i$ (i.e.~hydrodynamic interactions). The  tangent vector along the filament is defined as
\begin{equation}
\displaystyle\t\ui=\frac{\partial\r\ui}{\partial s}\cdot
\end{equation}
The hydrodynamic resistance coefficients for motion parallel and perpendicular to the local tangent,  $\xi_\parallel$ and $\xi_\perp$,   approximately satisfy
\begin{equation}
	\xi_\perp=2\xi_\parallel,
\end{equation}
which leads to a simpler form of Eq.~\ref{rft} as
\begin{equation}\label{rft_sim}
\f_h\ui= -\xi_\perp\left[\I-\frac{1}{2}\t\ui\t\ui\right]\cdot\left[\frac{\partial\r\ui}{\partial t}-\v\uji\right].
\end{equation}
Using the scalings consistent with Eq.~\ref{ND_scalings}, 
\begin{equation}\label{scale_v}
{\bar{\mathbf{f}}}_h=\frac{L^3}{A}\f_h,\quad \bv\uji=\v\uji\frac{\xi_\perp L^3}{A},
\end{equation}
the dimensionless form of Eq.~\ref{rft_sim} is finally given by
\begin{equation}\label{fh_nd}
	{\bar{\mathbf{f}}}_h\ui=-\left[\I-\frac{1}{2}\t\ui\t\ui\right]\cdot\left[\frac{\partial\br\ui}{\partial \bt}-\bv\uji\right].
\end{equation}

\subsection{Hydrodynamic interactions}
Hydrodynamic interactions between the moving filaments fundamentally arise from two different types of motion, which will be captured by different flow singularities:  rotational motion of the filaments around their centreline (fast decaying rotlets) and translational motion (slow decaying stokeslets). 
The dimensional flow field induced by the motion of filament $j$ near filament $i$,  $\v\uji$, can therefore be split into two terms induced by local moments and forces as
\begin{equation}\label{v_separate}
\v\uji=\v_M\uji+\v_F\uji.
\end{equation}
Due to the wide-separation assumption, $a\ll h_0$, both $\v_M$ and $\v_F$ can be described by a superposition of flow singularities.

Using $\mathbf{m}\uj$ to denote the torque density acting on filament $j$, and $\R(s, s')$ the relative position vector $\r\ui(s)-\r\uj(s')$, with magnitude $R(s, s')$, the flow $\v_M\uji$  is given by an integration of rotlets singularities \cite{Batchelor1970},
\begin{eqnarray}
\v_M\uji=\int_0^L\frac{-\mathbf{m}\uj(s')}{8\pi\mu}\times\frac{\R(s, s')}{R(s, s')^3} ds',
\end{eqnarray}
where the minus sign arises from the fact that $-\mathbf{m}\uj$ is the density of moment acted from the filament on the fluid. 

Using the classical hydrodynamic result
\begin{equation}
\mathbf{m}\uj=-\xi_r\omega\uj\t\uj,
\end{equation}
the integral becomes
\begin{eqnarray}
\v_M\uji=\int_0^L\frac{\xi_r\omega\uj}{8\pi\mu}\t\uj\times\frac{\R(s, s')}{R(s, s')^3} ds'.
\end{eqnarray}

Similarly, the flow velocity induced by the translational motion is given by a linear superposition of stokeslets as
\begin{eqnarray}
\v_F\uji(s)=\int_0^L\frac{-\f_h\uj(s')}{8\pi\mu}\cdot\left(\frac{\mathbf{I}}{R}+\frac{\R\R}{R^3}\right)ds',
\end{eqnarray}
where the force density along filament $j$, $\f_h\uj$, may be obtained from   Eqs.~\ref{FB} and ~\ref{fe_nd}.

Following the  nondimensionalisation procedures from  Eq.~\ref{ND_scalings} and \ref{scale_v}, the dimensionless form of  these two integrals are given by
\begin{subeqnarray}\label{eq:integrals}
\bar{\v}_M\uji(\bar{s})&=&\frac{\epsilon_a^2}{2}\int_0^1 \bar{\omega}(\bar{s}')\uj\t\uj(\bar{s}')\times\frac{\bar{\R}(\bar{s},\bar{s}')}{\bar{R}(\bar{s},\bar{s}')^3}d\bar{s}',\\
\bar{\v}_F\uji(\bar{s})&=&\frac{1}{2\ln(1/\epsilon_a)}\int_0^1{\bar{\mathbf{f}}}_h\uj\cdot\left(\frac{\I}{\bar{R}}+\frac{\bar{\R}\bar{\R}}{\bar{R}^3}\right)d\bar{s}'.
\end{subeqnarray}

\begin{figure}[t]
\centering
\includegraphics[width=0.6\textwidth]{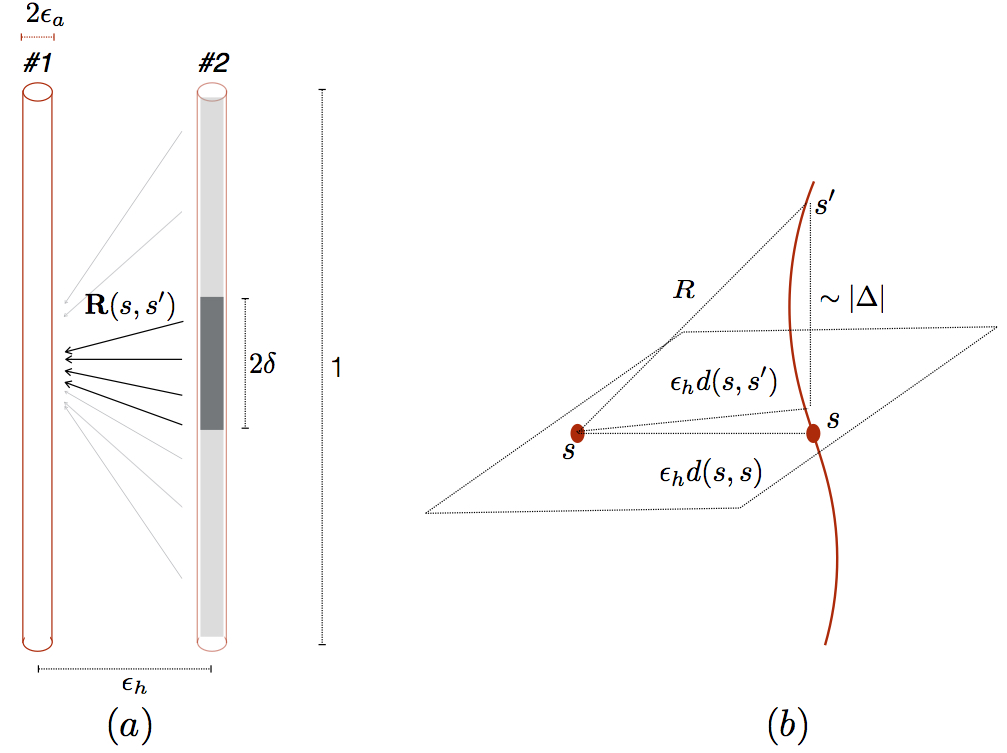}
\caption{
($a$) Illustration of the mathematical method used to compute hydrodynamic interactions. The integration domain  is divided into a local region of length $2\delta$ (dark gray) and non-local region (light gray); 
($b$) Geometric relations between $d(s,s')$ and $R$: $R$ is the distance between two points on the two filaments; $\epsilon_hd(s,s')$ is its  projection in $x-y$ plane; $\epsilon_hd(s,s)$ is the local separation distance between two points at the same arclength $s$.}\label{fig_scheme}
\end{figure}

Inspired by a classical calculation by Lighthill \cite{lighthill75}, the mathematical approach used to carry out the integrals in Eq.~\ref{eq:integrals} is illustrated in Fig.~\ref{fig_scheme}. We aim to compute the leading order term of $\v\uji$ asymptotically at limit $\epsilon_a\ll \epsilon_h\ll 1$. In order to carry out these evaluations  despite the fact that the kernels in Eq.~\ref{eq:integrals} are singular at $\bar R= 0$, we introduce an intermediate (arbitrary) length $\delta$ satisfying $\epsilon_h\ll\delta\ll 1$ in order  to divide the integration regions into local and non-local parts. In the local part, 
the filament is nearly straight which simplifies the algebra while in the non-local part the kernels are no longer singular and the integrals may be evaluated asymptotically directly. The final asymptotic results, sum of the two integration regions, should be independent of the value of $\delta$.

For notation convenience, we drop the bars below to indicate dimensionless quantities in the following derivations.

\subsection{Nonlocal term}
Introducing the notation $\Delta=s-s'$, the nonlocal region is where $|\Delta|\geqslant\delta$, and  the corresponding integral contributions to the velocities, denoted $NL$, are given by
\begin{subeqnarray}\label{vNL}
	\v_M^{NL}(s)&=&\frac{\epsilon_a^2}{2}\int_{s-1}^{-\delta}+\int_{\delta}^{s}\omega\uj(s-\Delta)\t\uj(s-\Delta)\times\frac{\R}{R^3}d\Delta,\\
\v_F^{NL}(s)&=&\frac{1}{2\ln(1/\epsilon_a)}\int_{s-1}^{-\delta}+\int_{\delta}^{s}\f_h\uj(s-\Delta)\cdot\left(\frac{\I}{R}+\frac{\R\R}{R^3}\right)d\Delta.
\end{subeqnarray}

As illustrated in Fig.~\ref{fig_scheme}b, we write the relative position vector as,
\begin{eqnarray}\label{expandNL}
	\R(s,s') &=&\Delta\e_z+\epsilon_h\mathbf{d}(s,s'),
\end{eqnarray}
with magnitude written $R=\sqrt{\Delta^2+\epsilon_h^2d^2(s,s')}$ and where the vector $\d(s,s')$ denotes the projection of $\R$ in $x-y$ plane at point $s$.  Since $\Delta\geqslant\delta\gg\epsilon_h$, we have $R(s,s')\sim|\Delta|$. Substituting this approximation into Eq.~\ref{vNL}, we have approximately
\begin{subeqnarray}
\slabel{vMNL}
\v_{M}^{{\rm NL}}&=&\frac{\epsilon_a^2}{2}\int_{s-1}^{-\delta}+\int_{\delta}^{s}\omega\uj(s-\Delta)\e_z\times\frac{\Delta\e_z+\epsilon_h\d(s,s-\Delta)}{|\Delta|^3}d\Delta\\
&=&\frac{\epsilon_a^2\epsilon_h}{2}\int_{-\delta}^{s-1}+\int_{\delta}^{s}\omega\uj(s-\Delta)\e_z\times\d(s,s-\Delta)\frac{d\Delta}{\Delta^3},\notag\\
\slabel{vFNL}
\v_F^{\rm NL}&=&\frac{1}{2\ln(1/\epsilon_a)}\int_{-\delta}^{s-1}+\int_{\delta}^{s}\f_h\uj(s-\Delta)\cdot\left(\frac{\I}{\Delta}+\frac{\e_z\e_z}{\Delta^3}\right)d\Delta.
\end{subeqnarray}

In the limit where $\delta\to 0$, the integrals in Eqs.~(\ref{vMNL}) and (\ref{vFNL})  diverge  and are dominated by the behaviour of the integrands near the boundary ($\Delta=0$). Simplifying the notations as
\begin{equation}
\d(s,s)\equiv \d,\quad d(s,s)\equiv d,	
\end{equation}
then we obtain 
\begin{subeqnarray}
\slabel{vMNL1}
\v_{M}^{{\rm NL}}&=&\frac{\epsilon_a^2\epsilon_h}{2}\omega\uj(s)\e_z\times\d\int_{s-1}^{-\delta}+\int_{\delta}^{s}\frac{d\Delta}{|\Delta|^3},\\
\slabel{vFNL1}
\v_F^{\rm NL}&=&\frac{1}{2\ln(1/\epsilon_a)}\f_h\uj(s)\cdot\int_{-\delta}^{s-1}+\int_{\delta}^{s}\left(\frac{\I}{|\Delta|}+\frac{\e_z\e_z}{|\Delta|^3}\right)d\Delta.
\end{subeqnarray}
Considering the integrations
\begin{subeqnarray}
	\int_{s-1}^{-\delta}+\int_{\delta}^{s}\frac{d\Delta}{|\Delta|^3}&=&\frac{1}{\delta^2}+O(1),\\
	\int_{s-1}^{-\delta}+\int_{\delta}^{s}\frac{d\Delta}{|\Delta|}&=&-2\ln\delta+O(1),
\end{subeqnarray}
then we obtain
\begin{subeqnarray}\label{vNL_result}
\v_M^{\rm NL}(s)&=&\frac{\epsilon_a^2\epsilon_h}{2\delta^2}\omega\uj(s)\e_z\times\d,\\
\v_F^{\rm NL}(s)&=&\frac{\ln\delta}{\ln\epsilon_a}\f\uj_h(s)\cdot\left(\I+\e_z\e_z\right),
\end{subeqnarray}
plus terms which are higher order in $\delta$.

\subsection{Local term}
Next we consider local integration where $|\Delta|\leq\delta$, and therefore for which the integrals are given by
\begin{subeqnarray}\label{vL}
	\v_M^{\rm L}(s)&=&\frac{\epsilon_a^2}{2}\int_{-\delta}^{\delta}\omega\uj(s-\Delta)\t\uj(s-\Delta)\times\frac{\R}{R^3}d\Delta,\\
\v_F^{\rm L}(s)&=&\frac{1}{2\ln(1/\epsilon_a)}\int_{-\delta}^{\delta}\f\uj_h(s-\Delta)\cdot\left(\frac{\I}{R}+\frac{\R\R}{R^3}\right)d\Delta.
\end{subeqnarray}
The relevant  quantities in both integrands can be Taylor-expanded near $\Delta=0$ as
\begin{align}
	\left(
	\begin{matrix}
	 \omega\uj(s-\Delta)\\ \d(s, s-\Delta)\\ \r\uj(s-\Delta)
	\end{matrix}
	\right)=
	\left(
	\begin{matrix}
		 \omega\uj(s)\\ \d\\ \r\uj(s)
    \end{matrix}
    \right)-\Delta
    \left(
    \begin{matrix}
    	 \omega\uj_{0\Delta}(s)\\ \d_{0\Delta}(s, s)\\ \r\uj_{0\Delta}(s)
    \end{matrix}
    \right)+O(\Delta^2).
\end{align}
Under the long wavelength assumption, all geometrical and kinematic quantities vary on the length scale of the filaments, $L$, and thus 
the terms $\omega\uj_{0\Delta}$, $\d_{0\Delta}$ and  $r\uj_{0\Delta}(s)$ are all of order one. 
The leading-order contributions in $\delta$  of the integrals in Eq.~\ref{vL} can thus be obtained by using the local approximation, writing
\begin{subeqnarray}\label{vL_im}
	\v_{M}^{{\rm L}}(s)&=&\frac{\epsilon_a^2\epsilon_h}{2}\omega\uj(s)\e_z\times\d\int_{-\delta}^{\delta}(\Delta^2+\epsilon_h^2d^2)^{-\frac{3}{2}}d\Delta,\\
	\v_F^{{\rm L}}(s)&=&\frac{1}{2\ln(1/\epsilon_a)}\int_{-\delta}^{\delta}\f_h\uj(s)\cdot\left(\frac{\I}{\sqrt{\Delta^2+\epsilon_h^2d^2}}+\frac{\Delta^2\e_z\e_z}{(\Delta^2+\epsilon_h^2d^2)^\frac{3}{2}}\right)d\Delta.
\end{subeqnarray}

The elementary integrals in Eq.~\ref{vL_im}  may all be evaluated analytically as\begin{subeqnarray}
\int_{-\delta}^{\delta}\frac{d\Delta}{\sqrt{\Delta^2+\epsilon_h^2d^2}}&=&-2\ln(\epsilon_hd)+2\ln\delta+O(1),\\
\int_{-\delta}^{\delta}\frac{d\Delta}{\left(\Delta^2+\epsilon_h^2d^2\right)^{\frac{3}{2}}}&=&\frac{2}{\epsilon_h^2d^2}-\frac{1}{\delta^2}+O(1),\\
\int_{-\delta}^{\delta}\frac{\Delta^2d\Delta}{\left(\Delta^2+\epsilon_h^2d^2\right)^{\frac{3}{2}}}&=&\int_{-\delta}^{\delta}\frac{d\Delta}{\sqrt{\Delta^2+\epsilon_h^2d^2}}-\epsilon_h^2d^2\int_{-\delta}^{\delta}\frac{d\Delta}{\left(\Delta^2+\epsilon_h^2d^2\right)^{\frac{3}{2}}}\cdot
\end{subeqnarray}
and we obtain asymptotically in the limit of small $\delta$
\begin{subeqnarray}\label{vL_result}
\v_M^{\rm L}&=&\frac{\epsilon_a^2\epsilon_h}{2}\left(\frac{2}{\epsilon_h^2d^2}-\frac{1}{\delta^2}\right)\omega\uj\e_z\times\d,\\
\v_F^{\rm L}&=&\frac{1}{\ln\epsilon_a}\ln\left(\frac{\epsilon_h d}{\delta}\right)\f_h\uj(s)\cdot\left(\I+\e_z\e_z\right).
\end{subeqnarray}

\subsection{Final asymptotic result}
Adding the nonlocal and local results of Eqs.~\ref{vNL_result} and \ref{vL_result} together,  we obtain results independent of the value of $\delta$ as
\begin{subeqnarray}\label{v_result}
\v_M\uji &=& \frac{\epsilon_a^2}{\epsilon_h d^2}\omega\uj(s)\e_z\times\mathbf{d},\\
\v_F\uji &=&\frac{\ln\left(\epsilon_hd\right) }{\ln\epsilon_a}\f_h\uj(s)\cdot\left(\I+\e_z\e_z\right).\slabel{b}
\end{subeqnarray}
The algebraic dependence of $\v_M$ on $d$ arises from the fast $\sim 1/r^3$ spatial decay of rotlets while the logarithmic dependence in $\v_F$ is a consequence of the slow $\sim1/r$ decay of point forces in Stokes flows.

\section{Long-wavelength bundling model}

With both the elastic and hydrodynamic forces evaluated asymptotically in the limit  relevant to the bundling of bacterial flagella ($\epsilon_a\ll \epsilon_h\ll 1$), we may now derive the long-wavelength bundling model.

First, we substitute Eq.~\ref{fe_nd} and Eq.~\ref{fh_nd} into Eq.~\ref{FB}, so that the force balance on filament $i$ may now be written
\begin{equation}\label{finalEq_1}
	\frac{\partial^4\r\ui}{\partial s^4}+\left[\I-\frac{1}{2}\t\ui\t\ui\right]\cdot\left[\frac{\partial\r\ui}{\partial t}-\v\uji\right]={\bf 0}.
\end{equation}
As seen in Sec.~\ref{sec:twist}, the twist density is quasi-steady  mechanical equilibrium, and as a consequence the local rotation rate of the  filament relative to the fluid is constant, i.e.~$\omega\ui(s)=\omega_0\ui$. Since at leading order  $\t\ui$ is $\e_z$, we have the force balance as
\begin{equation}\label{finalEq_2}
		\frac{\partial^4\r\ui}{\partial s^4}+\left[\I-\frac{1}{2}\e_z\e_z\right]\cdot\left[\frac{\partial\r\ui}{\partial t}-\v\uji\right]={\bf 0}.
\end{equation}
Considering the description of the geometry in Cartesian coordinates as in Eq.~\ref{position}, we see that 
\begin{equation}\label{eq60}
\e_z\cdot\frac{\partial\r\ui}{\partial t}=0.
\end{equation}
Also, considering Eqs.~\ref{v_separate} and \ref{v_result}, and using the fact that the force density in Eq.~\ref{b} is given by Eq.~\ref{fe_nd}, we obtain
\begin{equation}\label{eq61}
	\v\uji=\frac{\epsilon_a^2\omega_0\uj}{\epsilon_h d^2}\e_z\times\d-\frac{\ln(\epsilon_h d)}{\ln\epsilon_a}\frac{\partial^4\r\uj}{\partial s^4},
\end{equation}
which clearly leads to
\begin{equation}\label{eq62}
	\e_z\cdot\v\uji=0.	
\end{equation}
Substituting Eqs.~\ref{eq60} and \ref{eq62} into Eq.~\ref{finalEq_2}, we have
\begin{equation}\label{finalEq_3}
		\frac{\partial^4\r\ui}{\partial s^4}+\frac{\partial\r\ui}{\partial t}=\v\uji,
\end{equation}
which, when using Eq.~\ref{eq61}, transforms into
\begin{align}\label{finalEq}
\frac{\partial\r\ui}{\partial t}	+\frac{\partial^4\r\ui}{\partial s^4}=\frac{\epsilon_a^2\omega_0\uj}{\epsilon_h d^2}\e_z\times\d-\frac{\ln(\epsilon_h d)}{\ln\epsilon_a}\frac{\partial^4\r\uj}{\partial s^4}\cdot
\end{align}

An inspection of the terms in Eq.~\ref{finalEq} reveals the presence of two dimensionless groups. The first one, which is multiplying the bending term, is given by a ratio of logarithms 
\begin{equation}
{\cal L}\equiv\frac{\ln(\epsilon_h d)}{\ln\epsilon_a}\cdot
\end{equation} 
Technically, this term is not a correct dimensionless group since it depends on the local filament-filament distance, $d$. However, given the  presence of two logarithms, it is expected to show only weak  dependence on it, and is thus approximately given by the ratio \cite{goldstein16}
\begin{equation}
{\cal L}_0\equiv\frac{\ln(L/h_0)}{\ln (L/a)}\cdot
\end{equation}

The second dimensionless number, more important,  controls the driving force in the bundling dynamics. It  is the one comparing the rotational component of the flow in Eq.~\ref{finalEq} to the bending (or viscous terms) which  scales as $\epsilon_h$. We term this dimensionless ratio  the {\it  Bundling number}, $\rm Bu$,  which is therefore defined by
\begin{equation}\label{eq:Bu}
{\rm Bu}\equiv\frac{\epsilon_a^2\omega_0}{\epsilon_h^2},
\end{equation}
where the extra factor of $\epsilon_h$ arises due to the scaling of $\r$. Clearly,  if different filaments have different rotation rates then  there will be more than one relevant Bundling number characterising the system. 
Alternatively, given Eq.~\ref{Sp}, the Bundling number can also be written using the $\rm Sp$ number
\begin{equation}
	{\rm Bu}=\frac{\epsilon_a^2{\rm Sp}^4}{\epsilon_h^2},
\end{equation}
and is given using dimensional quantities as
\begin{equation}\label{Bu-dimensional}
	{\rm Bu}= \frac{\xi_\perp\omega_0  a^2 L^4}{h_0^2A},
\end{equation}
where $\omega_0$ in Eq.~\ref{Bu-dimensional} refers to its dimensional form.  As a final note, we may write the final dimensionless model projected along Cartesian coordinates, and obtain
\begin{align}\label{finalEq_components}
\frac{\partial}{\partial t}
\begin{bmatrix}
	x\ui\\ y\ui
\end{bmatrix}
	+\frac{\partial^4}{\partial s^4}
	\begin{bmatrix}
		x\ui\\ y\ui 
	\end{bmatrix}
   =\frac{\epsilon_a^2\omega\uj_0}{\epsilon_h^2d^2}
    \begin{bmatrix}
    	-y\ui+y\uj\\
    	x\ui-x\uj
    \end{bmatrix}-\frac{\ln(\epsilon_h d)}{\ln\epsilon_a}
    \frac{\partial^4}{\partial s^4}
    \begin{bmatrix}
    x\uj\\y\uj	
    \end{bmatrix},
\end{align}
where $d^2={(x\ua-x\ub)^2+(y\ua-y\ub)^2}$.
This is the form which will be used below in our numerical solution of the problem.

\section{Bundling and unbundling of elastic filaments}

We have derived above the local partial differential equation describing the time-evolution of the elastic filaments  interacting hydrodynamically. It is given in Eq.~\ref{finalEq} in a vector form and in Eq.~\ref{finalEq_components} in Cartesian coordinates. We now study numerically (with technical details summarised in \S\ref{sec:numerical}) the dynamics predicted by the model and focus on four situations of interest: Two filaments with identical rotation rate (\S\ref{sec:twofilaments});  $N>2$ symmetric filaments (\S\ref{sec:N});  An asymmetric  of arrangement of three filaments (\S\ref{sec:asymm}) and the unbundling of filaments (\S\ref{sec:unbundling}). We will next analyse  the conformation instabilities observed numerically in  
\S\ref{sec:inst}.

\subsection{Numerical procedure}
\label{sec:numerical}
In order to solve Eq.~\ref{finalEq}, we discretise the equation spatially using second-order central finite  difference. Each filament is assumed to be  clamped at its rotated  end $(s=0)$ and free on the other end  ($s=1$) and thus the boundary conditions are written as
\begin{subeqnarray}\label{BCs}
\r(s=0,t)&=&\r_0,\quad \frac{\partial \r}{\partial s}(s=0,t)=\e_z,\\
\frac{\partial^2\r}{\partial s^2}(s=1,t)&=&\mathbf{0}, \quad \frac{\partial^3\r}{\partial s^3}(s=1,t)=\mathbf{0},
\end{subeqnarray}
where $\r_0$ refers to the location of the end point of  the particular filament of interest. 
Each filament starts initially from  a straight shape parameterised by
\begin{equation}
\r(s,t=0)=\r_0+s\e_z,\quad 0 \leq s \leq 1,
\end{equation}
from which Eq.~\ref{finalEq} is solved in time using a second-order Crank-Nicolson method. We use the values $\epsilon_a=0.001$ and $\epsilon_h=0.1$ throughout the paper for the two shape dimensionless numbers.

In order to  prevent the interacting filaments from touching each other and overlap, we apply numerically a short-range repulsive potential of the form
\begin{equation}
V_r=\left(\frac{2\epsilon_a}{\epsilon_h d}\right)^{q}.
\end{equation}
In the results shown in the following sections  we choose the value $q=4$ so that this repulsion only comes into play when  $d\sim  a$ and thus does not affect any of the long-ranged hydrodynamic features. We also tried $q=6$  with no qualitative impact on the dynamics. 
In practice the minimum distance between the filaments in our simulations never goes below three filament radius, so lubrication stresses are not expected to play any role. Note also that we assume that the drag coefficients remain constant  even at these short inter-filament distances; although this is clearly a simplifying assumption, we do not expect the results to strongly depend on it.

\subsection{Bundling of two filaments}
\label{sec:twofilaments}
We first  consider the case of two filaments with identical rotation rate, $\omega_0$.  If we denote the projection of $\r$ on the $x-y$ plane as
$\tilde{\r}=\r\cdot(\I-\e_z\e_z),	
$ then 
 by symmetry  we  clearly have
\begin{equation}
\tilde{\r}\ua(s,t)+\tilde{\r}\ub(s,t)=\mathbf{0}.
\end{equation}
Or in components,
\begin{align}
	\begin{bmatrix}
	x\ua\\y\ua
	\end{bmatrix}
    +
    \begin{bmatrix}
    x\ub\\ y\ub
    \end{bmatrix}=\mathbf{0}.
\end{align}
Substituting this symmetry property into Eq.~\ref{finalEq_components}, we obtain that each  filament shape  satisfies
\begin{equation}
\frac{\partial\tilde{\r}}{\partial t}+\left[1-\frac{\ln(\epsilon_h d)}{\ln\epsilon_a}\right]\frac{\partial^4	\tilde{\r}}{\partial s^4}=\frac{2\Bu}{d^2}\e_z\times\tilde{\r},
\end{equation}
written in components as
\begin{align}\label{2filaments_components}
\frac{\partial}{\partial t}
\begin{bmatrix}
	x\\ y
\end{bmatrix}
	+\left[1-\frac{\ln(\epsilon_h d)}{\ln\epsilon_a}\right]\frac{\partial^4}{\partial s^4}
	\begin{bmatrix}
		x\\ y 
	\end{bmatrix}
   =\frac{2{\rm Bu}}{d^2}
    \begin{bmatrix}
    	-y\\
    	x
    \end{bmatrix},\quad 
    d^2= 4(x^2+y^2).
\end{align}

\begin{figure}[t]
\centering
\includegraphics[width=0.65\textwidth]{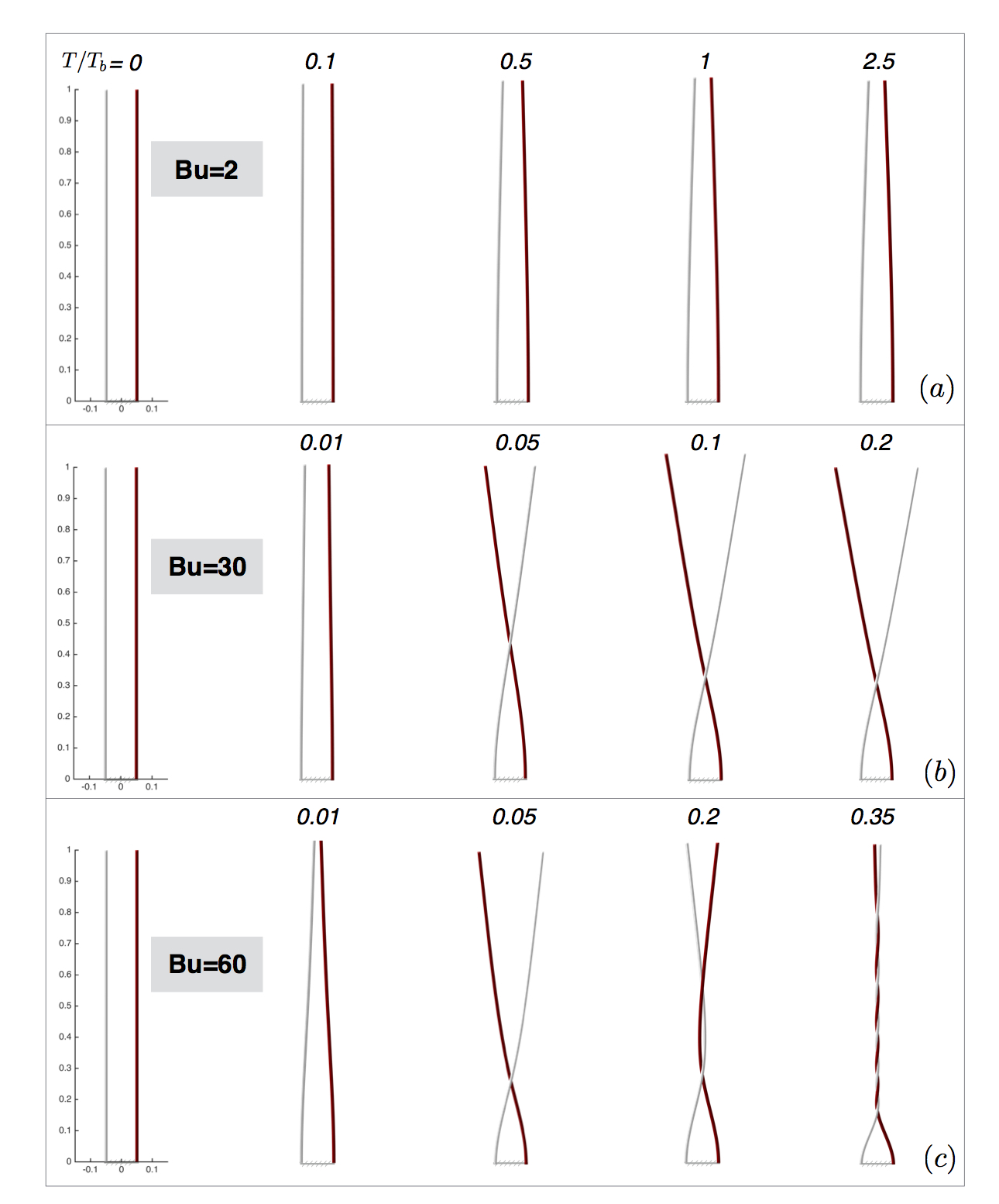}
\caption{Bundling of two elastic filaments as a function of time for three representative values of $\Bu$; ($a$): $\Bu=2$, with a weakly bent final configuration; ($b$): $\Bu=30$, with a final crossing configuration; ($c$): $\Bu=60$, with a final bundled state. The  bending energies corresponding to these shapes are plotted in Fig.~\ref{BE_t}. The dynamics  illustrated in this figure can be visualised in the movies available  in supplementary material  \cite{supp_yi}.
}\label{shape_t}
\end{figure}

\begin{figure}[t]
\centering
\includegraphics[width=0.6\textwidth]{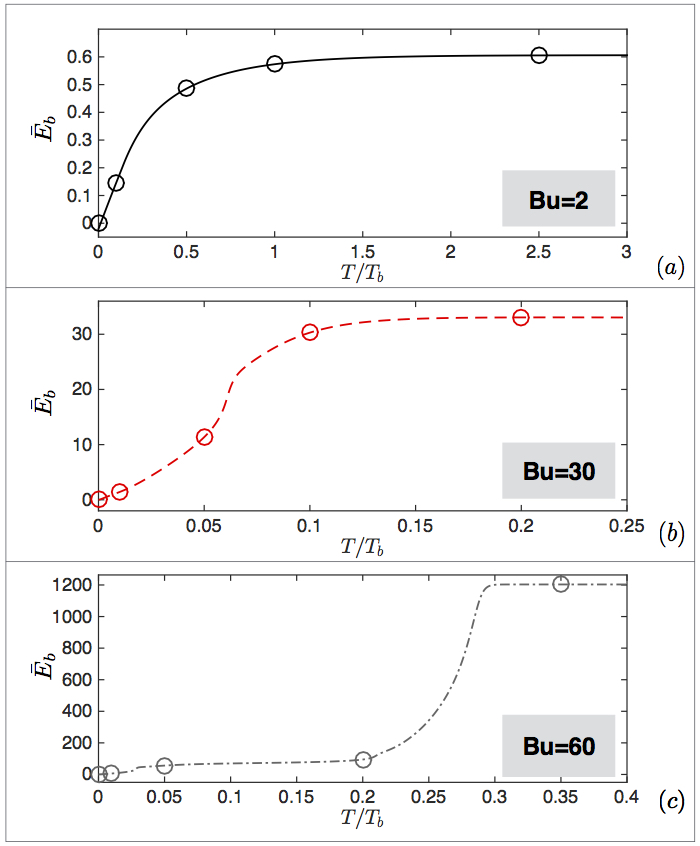}
\caption{Dimensionless filament bending energy for each filament, $\bar{E}_b$,  as a function of time for the dynamics illustrated in Fig.~\ref{shape_t} with specific times shown in circles: ($a$): $\Bu=2$;  ($b$): $\Bu=30$;     ($c$): $\Bu=60$. Here also the dynamics  illustrated in this figure can be visualised in the movies available  in supplementary material  \cite{supp_yi}.
}\label{BE_t}
\end{figure}

This equation was solved numerically  for a wide range of values of $\Bu$. As will be detailed below, we obtain three qualitatively different dynamics which we now illustrate by focusing on three representative values of $\Bu$ (specifically, $\Bu=2$, 30 and 60).  The time-varying shapes of the  filaments in these cases are shown as a function of time in Fig.~\ref{shape_t} while in Fig.~\ref{BE_t} we plot the dimensionless bending energy of each filament as a function of time. The entire dynamics shown in Figs.~\ref{shape_t}-\ref{BE_t} can also be visualised in the movies available  in supplementary material  \cite{supp_yi}.

The bending energy  for each filament, $E_b$, is equal to the  bending modulus times the  integral of curvature  square, i.e.
 \begin{equation}
 	E_b=\int_0^L \frac{A}{2}|\r_{ss}| ^2 ds.
 \end{equation}
 Non-dimensionalising as above, we may write 
 \begin{equation}
E_b=	\frac{A}{2L}\epsilon_h^2\bar{E}_b,
\end{equation}
where the dimensionless bending energy, $\bar{E}_b$, is given by
\begin{equation}
	 \bar{E}_b=\int_0^1( x_{\bs\bs}^2+y_{\bs\bs}^2) d\bar{s}.
\end{equation}

\begin{figure}[t]
\centering
\includegraphics[width=0.8\textwidth]{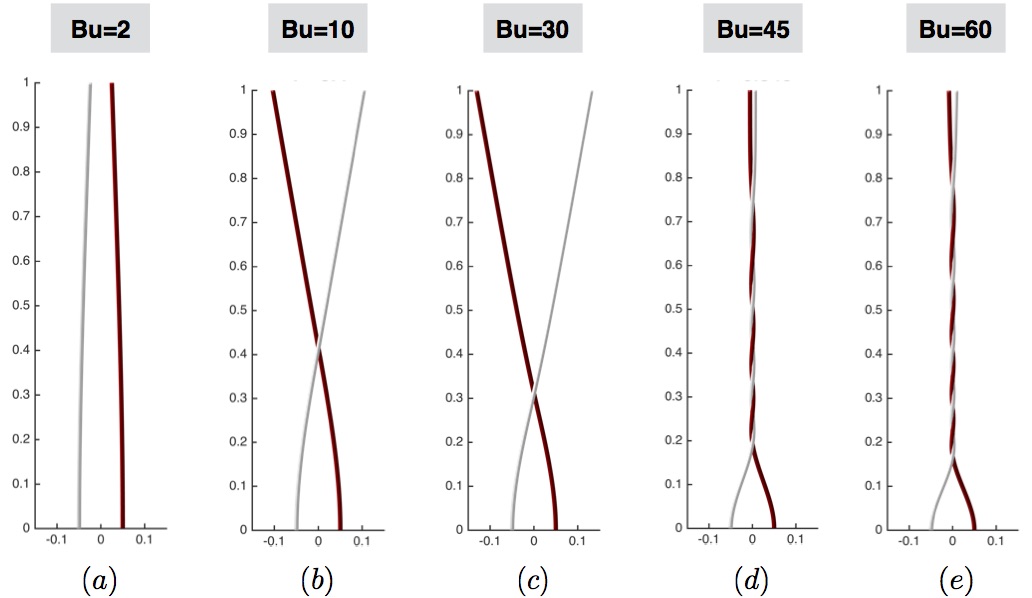}
\caption{Steady-state filament shapes for $\Bu=2$, $10$, $30$, $45$ and $60$ obtained  in the long-time limit. The three possible final states are: ($a$): weakly bent;  ($b$) and ($c$): crossing;  ($d$) and ($e$):  bundled.}\label{SteadyStates}
\end{figure}

 In all three cases the long-time steady-state shapes are illustrated in Fig.~\ref{SteadyStates}. For a small value, $\Bu<2$, the rotating filaments each bend toward each other until converging on steady weakly bent shapes (Fig.~\ref{shape_t}-\ref{SteadyStates}a). For an intermediate value, $\Bu=30$, the initial dynamics is similar but then at a critical time the filaments undergo a rapid conformation change and snap into  a crossing configuration in which they remain (Fig.~\ref{shape_t}-\ref{SteadyStates}b). In contrast for a high value, $\Bu=60$, the filaments  do  not remain in a crossing configuration but instead they quickly transition to a bundled state where the filaments wrap helically around each other  along the portion of the filament located after the crossing point
 (Fig.~\ref{shape_t}-\ref{SteadyStates}c)

\begin{figure}[t]
\centering
\includegraphics[width=0.7\textwidth]{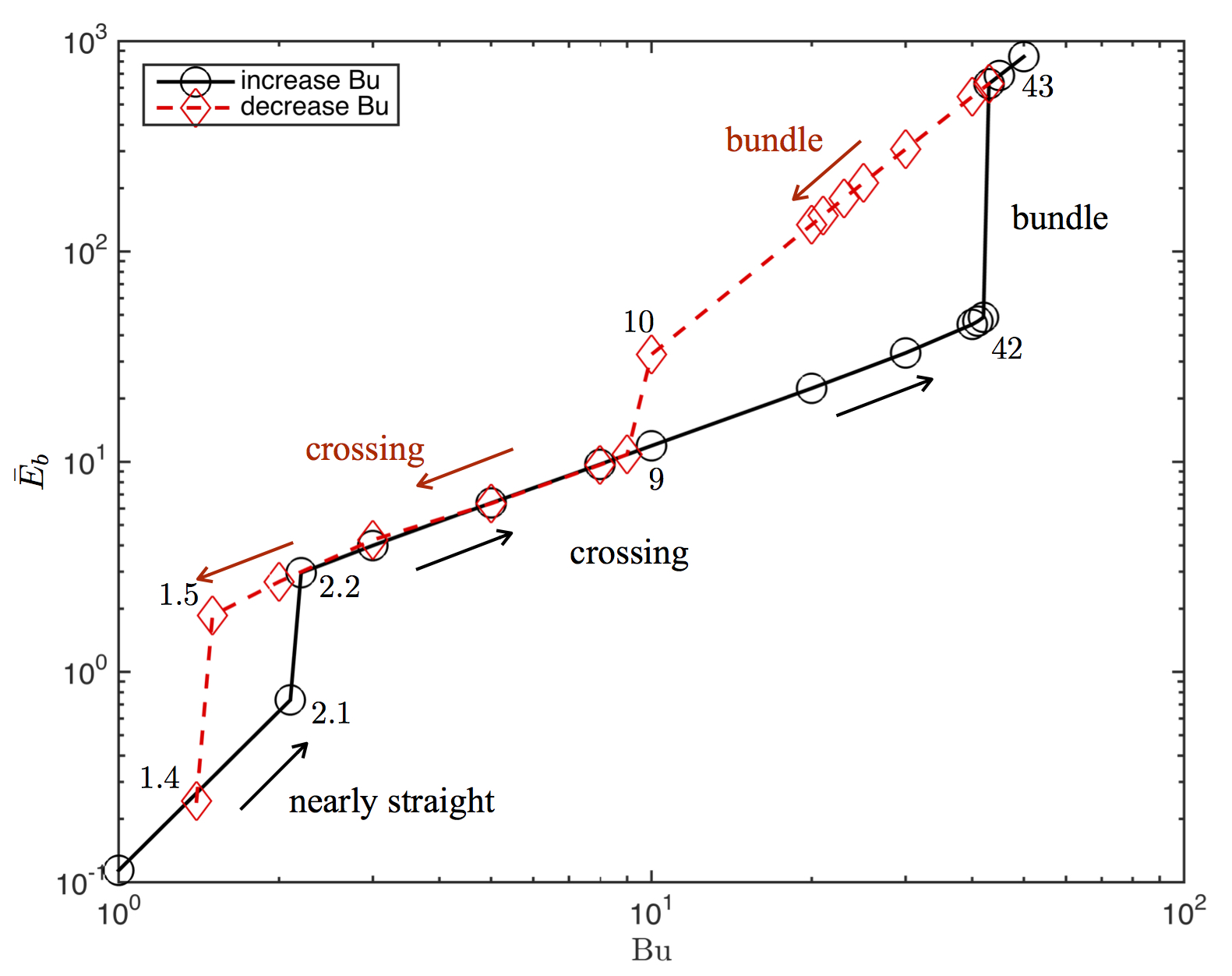}
\caption{Dimensionless bending energy for each filament, $	 \bar{E}_b$,  as a function of the Bundling number,  $\rm Bu$. The black solid line displays the steady-states when increasing the values of $\Bu$ while the  red dashed line shows the states obtained when decreasing  $\Bu$, showing two   hysteresis loops. The critical $\Bu$ values at the transitions in the increasing branch are $\Bu\approx2.1$ and 42, while for the decreasing branch they are $\Bu\approx10$ and 1.5. }\label{hysteresis}
\end{figure}

In order to further characterise the different steady-state conformations, we  plot in Fig.~\ref{hysteresis}  the value of the long-time dimensionless  bending energy of the filaments as a function of $\Bu$.  The results show two branches; the black solid line represents numerically results obtained while increasing the values of $\Bu$ while the red dashed line captures the computational results obtained by starting with a high value and decreasing $\Bu$. Interestingly, we see that the two transitions (bent $\leftrightarrow$ crossing and crossing $\leftrightarrow$ bundle) both show hysteresis loops.  When $\Bu$ increases from zero, the transitions occur approximately at $\Bu\approx 2.1$ and 42, while in the  decreasing case, the critical values are $\Bu\approx 10$ and 1.5.

\subsection{$N$ symmetric filaments}
\label{sec:N}

\begin{figure}[t]
\centering
\includegraphics[width=0.6\textwidth]{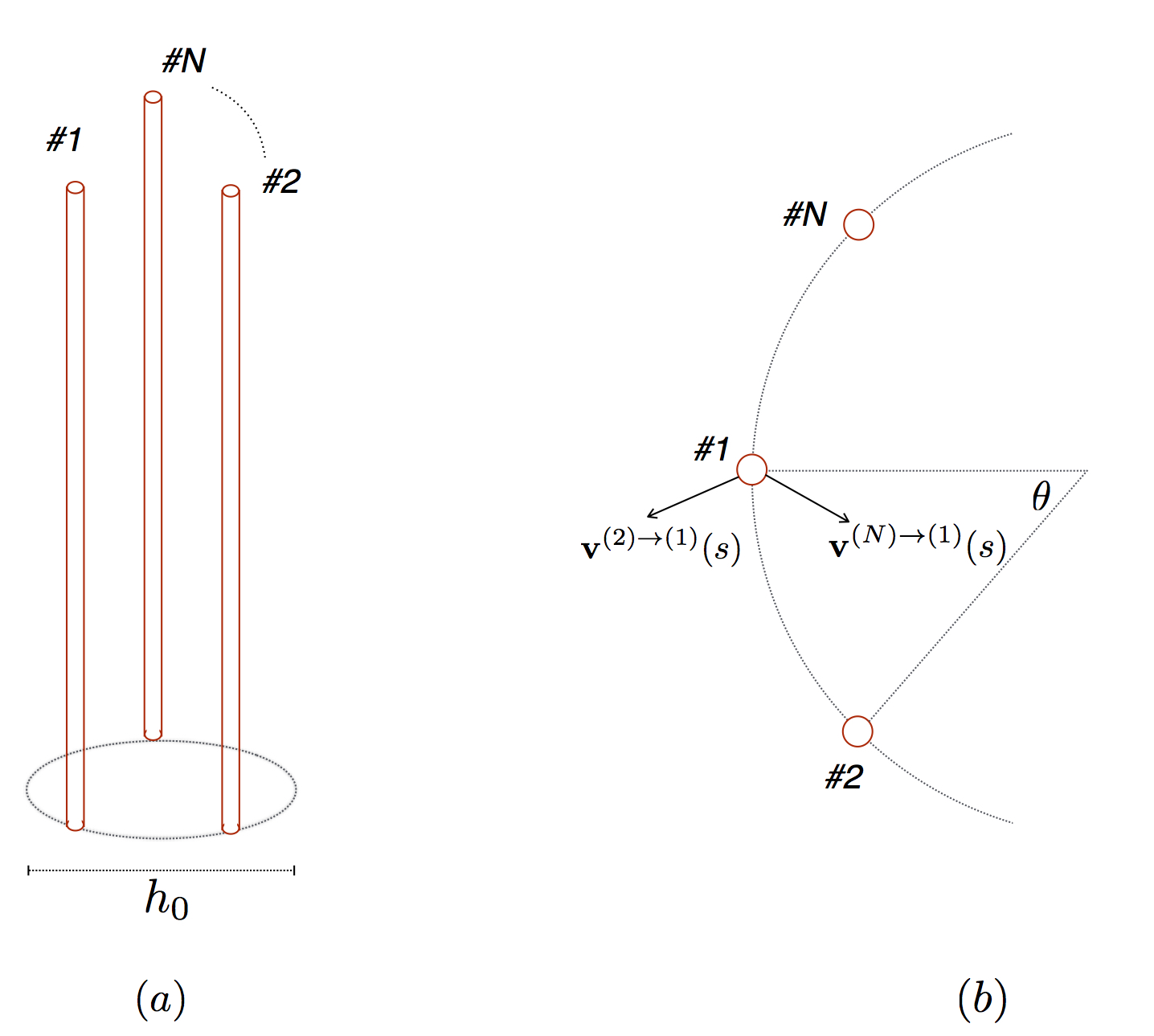}
\caption{$(a)$ Configuration of $N$ filaments symmetrically distributed along a circle of diameter $h_0$;  ($b$): Top view of the $N$ filament configuration; $\v\uji$ denotes the velocity generated by $j$th filament near the  $i$th one while $\theta=2\pi/N$. By exploiting the symmetries of the setup, the net flow generated by $j$th ($j=2,3...N$) and $(N+2-j)$th filaments is independent of $j$.
}\label{Multi}
\end{figure}

After focusing on two filaments, we can consider the more general case of $N$ filaments distributed symmetrically along a circular base from which they are all rotated with equal angular velocity. This situation is sketched in Fig.~\ref{Multi}.

The in-place location of filament \#$n$ is written as
\begin{equation}
\r\um(s)=[h_0x\um(s), h_0y\um(s)],
\end{equation}
which maybe be related to the  displacements $[\zeta\um,  \eta\um]$ as
\begin{equation}
x\um=\frac{\sin(n-1)\theta}{2}+\zeta\um,\quad y\um=-\frac{\cos(n-1)\theta}{2}+\eta\um.
\end{equation}
In order to simplify the system we define  the complex number
\begin{equation}
 \mathbf{z}\um=x\um+iy\um,
 \end{equation}
 which, due to the expected symmetry of the $N$-filament configuration,  satisfies
\begin{equation}
\mathbf{z}\um=\mathbf{z} e^{i(n-1)\theta}.	
\end{equation}

In order to derive the dynamics equation, we first  extend Eq.~\ref{finalEq} to the case of $N$ arbitrary filaments by adding up the flows arising from hydrodynamic interactions. Using dimensional quantities,  the position of the $nth$ filament satisfies
\begin{equation}\label{Multi_general}
\frac{\partial\r\um}{\partial t}	+\frac{A}{\xi_\perp}\frac{\partial^4\r\um}{\partial s^4}=a^2\omega_0\sum_{j\neq n}^{N}\frac{\e_z\times(\r\um-\r\uj)}{|\r\um-\r\uj|^2}-\frac{A}{\xi_\perp}\sum_{j\neq n}^{N}\frac{\ln(|\r\um-\r\uj|/L)}{\ln(a/L)}\frac{\partial^4\r\uj}{\partial s^4}\cdot
\end{equation}
The two summations in Eq.~\ref{Multi_general} maybe be expressed explicitly by exploiting symmetries.  Focusing on the filament with $n=1$, for which we write $\r^{(1)}\equiv\r$, we observe that
\begin{align}
\begin{split}
\sum_{l=j, N+2-j}\frac{\z-\z^{(l)}}{|\z-\z^{(l)}|^2}&=\frac{\z\left[2-e^{i(j-1)\theta}-e^{i(N+1-j)\theta}\right]}{|\z|^2\left[(1-\cos(j-1)\theta)^2+\sin^2(j-1)\theta\right]}=\frac{\z}{|\z|^2},
\end{split}
\end{align}
which is independent of $j$. As a result we have a first summation given by
\begin{equation}\label{sum1}
	\sum_{j=2}^{N}\frac{\e_z\times(\r-\r\uj)}{|\r-\r\uj|^2}=(N-1)\frac{\e_z\times\r}{2|\r|^2}\cdot
\end{equation}
Similarly for the second summation we have
\begin{align}
\begin{split}
	\sum_{l=j, N+2-j}\ln|\z-\z^{(l)}|\frac{\partial^4\z^{(l)}}{\partial s^4}=2\cos(j-1)\theta\left[\ln|\z|+\ln\sqrt{2-2\cos(j-1)\theta}\right]\frac{\partial^4\z}{\partial s^4},
\end{split}
\end{align}
so that
\begin{align}\label{sum2}
\begin{split}
\sum_{j=2}^{N}\ln|\r-\r\uj|\frac{\partial^4\r\rm}{\partial s^4}&=\frac{\partial^4\r}{\partial s^4}\sum_{j=2}^{N}\cos(j-1)\theta\left[\ln|\r|+\ln\sqrt{2-2\cos(j-1)\theta}\right]\\
&=\left(\alpha_N\ln|\r|+\beta_N\right)\frac{\partial^4\r}{\partial s^4},
\end{split}
\end{align}
where the coefficients $\alpha_N$ and $ \beta_N$ are given by
\begin{equation}
\alpha_N=\sum_{j=2}^{N}\cos(j-1)\theta=-1,\quad \beta_N=\sum_{j=2}^{N}\cos(j-1)\theta\ln\sqrt{2-2\cos(j-1)\theta}.
\end{equation}

\begin{figure}[t]
\centering
\includegraphics[width=0.7\textwidth]{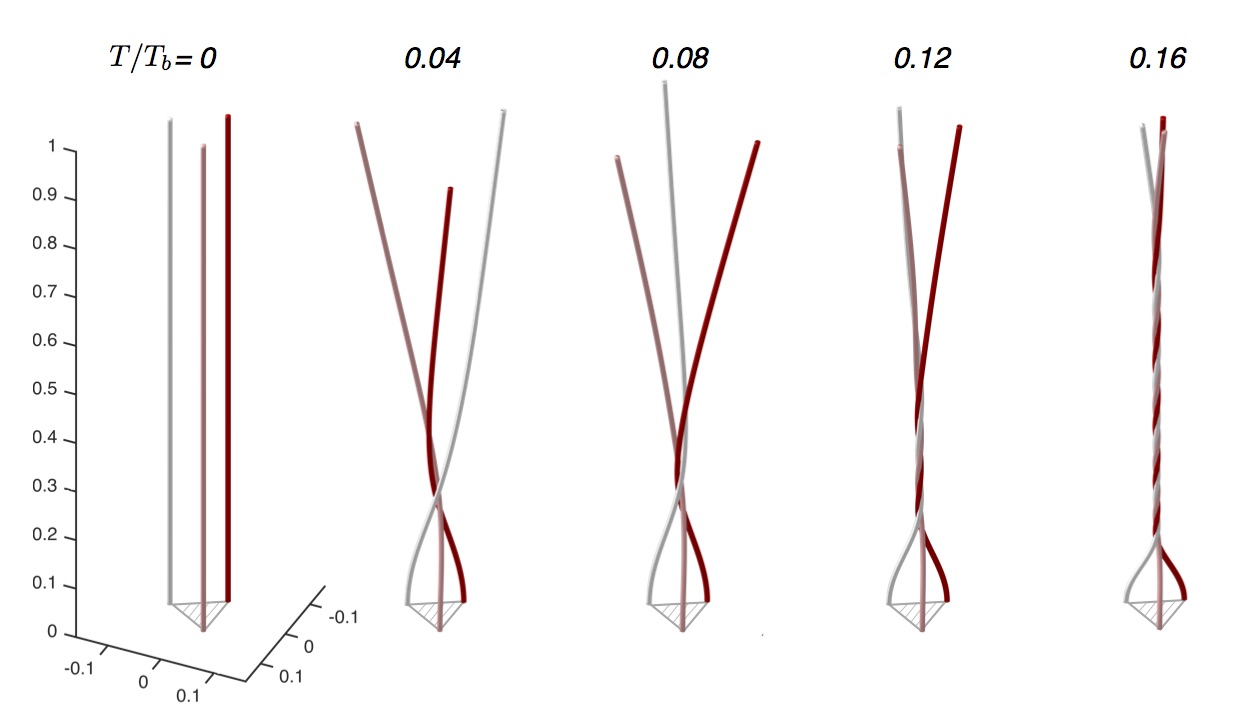}
\includegraphics[width=0.7\textwidth]{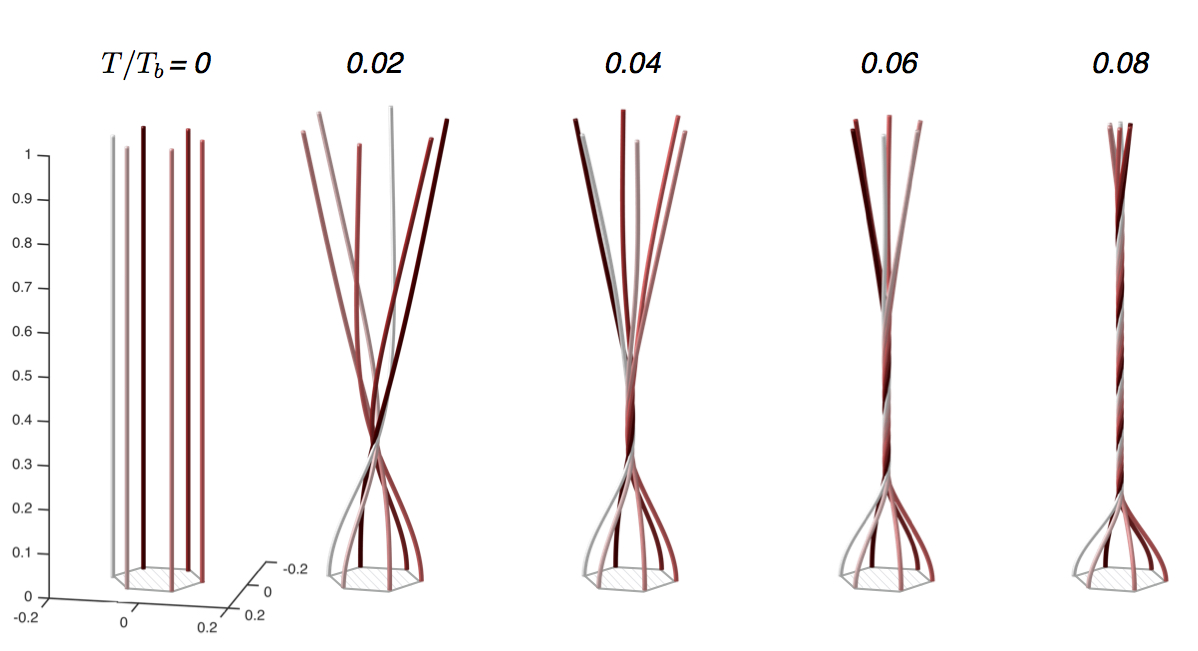}
\caption{Time dynamics of a symmetric arrangement of $N=3$ filaments (top) and $N=6$  (bottom) for $\Bu=60$ showing  crossing at intermediate times followed by a transition to a bundled state at long times. }
\label{NFilaments_SS3}
\label{NFilaments_SS6}
\end{figure}

Substituting Eq.~\ref{sum1} and \ref{sum2} into Eq.~\ref{Multi_general}, we obtained the simplified equation in the case of $N$ symmetric filaments
\begin{equation}
\frac{\partial\r}{\partial t}	+\frac{A}{\xi_\perp}\frac{\partial^4\r}{\partial s^4}=a^2\omega_0\frac{(N-1)\e_z\times\r\ua}{2|\r|^2}-\frac{A}{\xi_\perp\ln(a/L)}\left(\beta_N-\ln|\r/L|\right)\frac{\partial^4\r}{\partial s^4},
\end{equation}
which may also be written using   dimensionless variables as
\begin{equation}\label{eq:finalN}
\frac{\partial\br}{\partial \bar{t}}+\left[1+\frac{\left(\beta_N-\ln|\br|\right)}{\ln\epsilon_a}\right]\frac{\partial^4\br}{\partial \bs^4}=\frac{{\epsilon_a^2\bar{\omega}_0}(N-1)}{2}\frac{\e_z\times\br}{|\br|^2},
\end{equation}
where a $N$-filament Bundling number is naturally defined as
\begin{equation}
\Bu=	\frac{{\epsilon_a^2\bar{\omega}_0}(N-1)}{\epsilon_h^2}\cdot
\end{equation}
Note that the case  $N=2$ which was considered earlier in the paper may be recovered  under this framework. Indeed we have in this case
\begin{equation}
\theta=\frac{\pi}{2},\quad \beta_2=-\ln2,
\end{equation}
leading to
\begin{equation}
\frac{\partial\br}{\partial \bar{t}}+\left[1-\frac{\ln\left(2|\br|\right)}{\ln\epsilon_a}\right]\frac{\partial^4\br}{\partial \bs^4}=\frac{{\epsilon_a^2\bar{\omega}_0}}{2}\frac{\e_z\times\br}{|\br|^2}\cdot
\end{equation}
Since $	2\br=\epsilon_h \d$, the equation is identical to  Eq.~\ref{2filaments_components}.

The final equation, Eq.~\ref{eq:finalN}, allows thus to exploit symmetries in order to significantly simplify the complicated multi-filaments problem. Qualitatively, we obtain the  same three regimes (weakly bent, crossing, bundling) as with two filaments. Numerical results showing the time dependence of the shapes are illustrated in Fig.~\ref{NFilaments_SS3}  for $N=3$ filaments (top) and $N=6$ (bottom) in the case of a large Bundling number, $\Bu=60$, leading to crossing at intermediate times and a final bundled state.

\begin{figure}[t]
\centering
\includegraphics[width=0.75\textwidth]{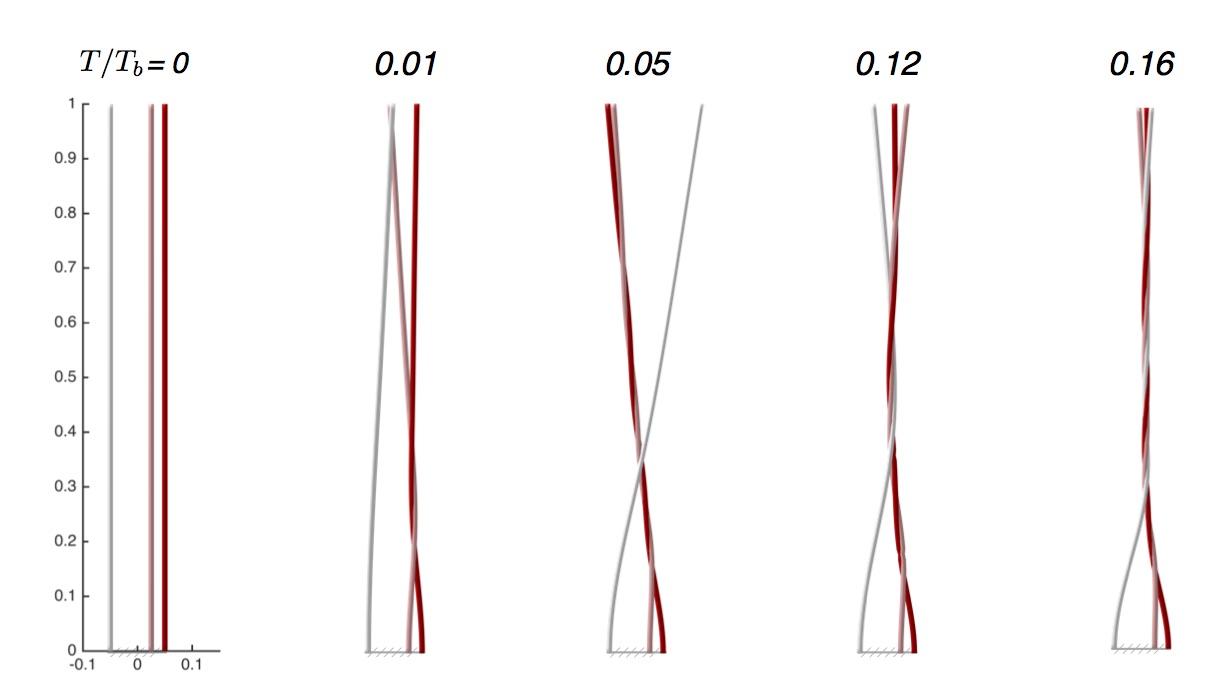}
\caption{Asymmetric bundling of three filaments clamped and rotated with a 3:1 ratio in their relative separation for $\Bu=30$.}\label{Asymmetry}
\end{figure}

\subsection{Asymmetric filaments}
\label{sec:asymm}

In the situation  where the distribution of filaments is not symmetric, it is possible to apply our model by solving the coupled system of equations in Eq.~\ref{Multi_general}. As an illustration, we show in Fig.~\ref{Asymmetry} the dynamics in the case of $N=3$ filaments clamped and rotated with a 3:1 separation distance along their base. The initial position of the filaments is  given in a dimensional form by
\begin{equation}
\r_0^{(1)}=[0, -0.05L],\quad \r_0^{(2)}=[0, 0.25L],\quad \r_0^{(3)}=[0, 0.05L].
\end{equation}
For $\Bu=30$ ($\epsilon_h$ is chosen as the largest ratio between filament-filament distance and $L$, which is 0.1 in this case), the two filaments which are the closest to each other bundle first together, and then bundle as a pair with the third filament located further away.

\begin{figure}[t]
\centering
\includegraphics[width=0.7\textwidth]{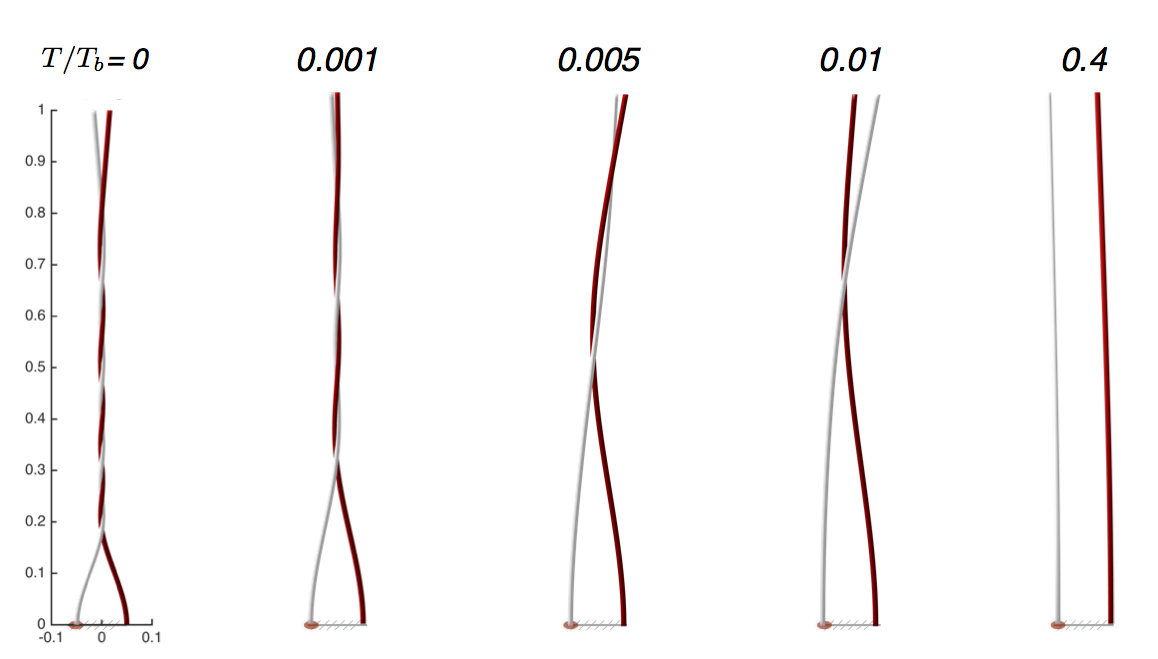}
\includegraphics[width=0.7\textwidth]{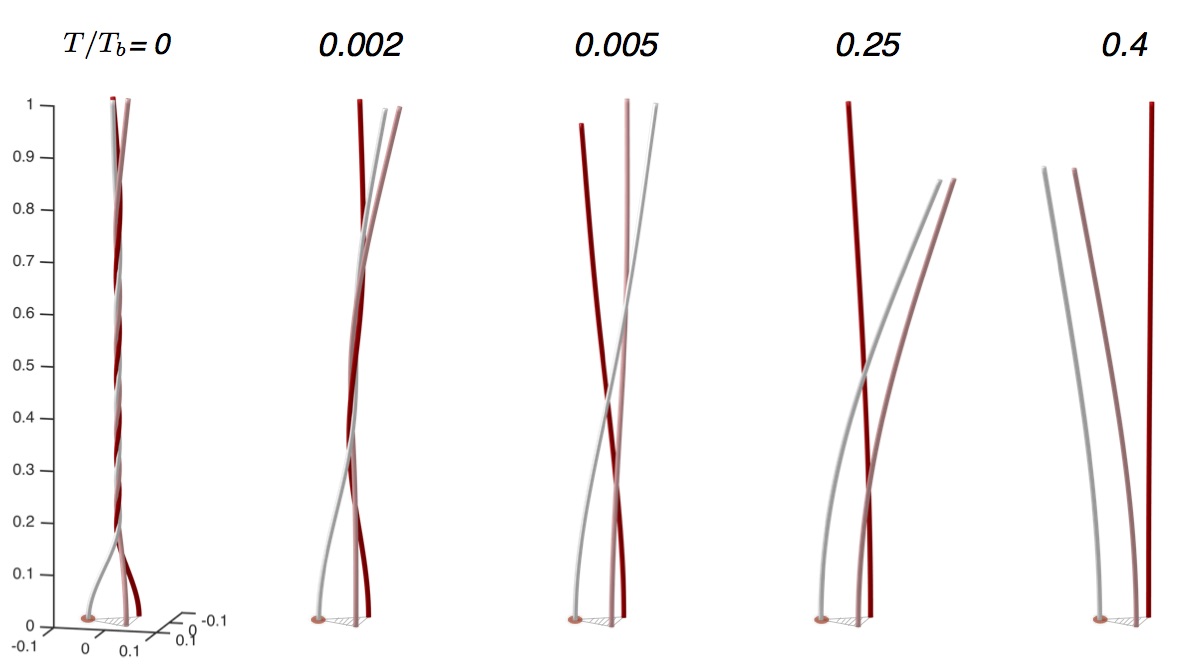}
\caption{Hydrodynamic unbundling of  rotating filaments. 
Top: cases of two bundled filaments where at $t=0$ the 
Bundling number of the filament on the left (with red dot at bottom) is switched to $\Bu=-50$, while for the other is  maintained at $\Bu=50$. Bottom: three filaments where the  filament with the red dot at bottom is switched to $\Bu= -40$, while for the other two are maintained at $\Bu= 40$. }
\label{UnBu2}
\label{UnBu3}
\end{figure}

\subsection{Unbundling}
\label{sec:unbundling}

\begin{figure}[t]
\centering
\includegraphics[width=0.7\textwidth]{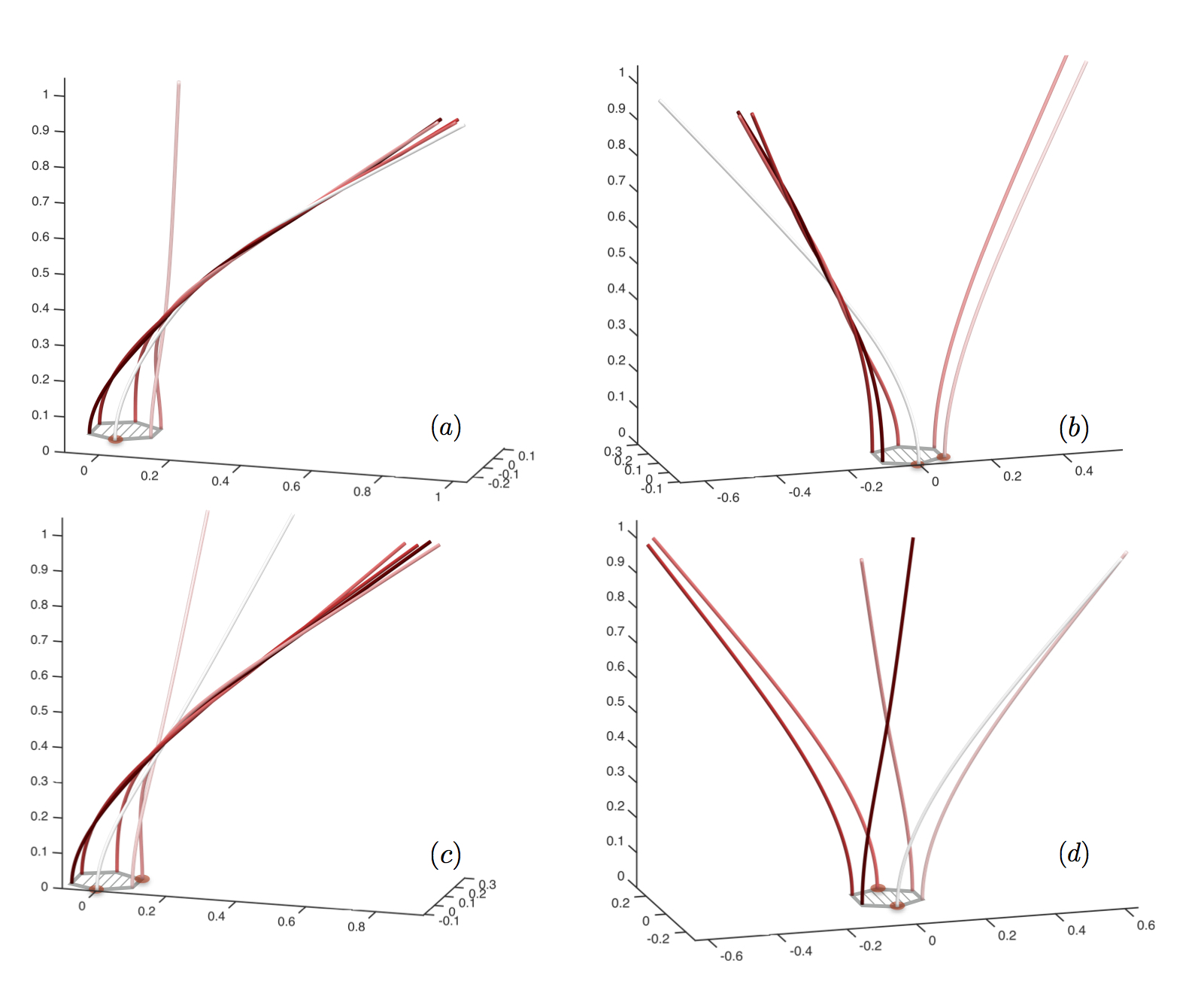}
\caption{Steady unbundled states for $N=6$ filaments.  Filaments with a red dot at their base have negative rotation and $\Bu=-40$ while all other ones are maintained at  $\Bu=40$. Counter-rotation is implemented for: (a) one filament; (b) two adjacent filaments; (c) two every-other filaments; (d) two opposite filaments.
}\label{UnBu6}
\end{figure}

In  the biological world, swimming bacteria change their swimming direction by reversing the rotation direction of at least one its rotary motors, leading the bundle of helical flagella to come apart  \cite{Turner00}. This process may be addressed with our simple model. We consider $N$ identically-rotating  filaments in the steady bundled state and may then switch the rotation direction of one of the filaments (i.e.~switching its Bundling number to a negative value). This is illustrated in Fig.~\ref{UnBu2} for $N=2$ filaments (top) and $N=3$ (bottom). In both cases, the unbundling process is fast compared to the initial bundling dynamics shown in Figs.~\ref{shape_t} and \ref{NFilaments_SS3}  and it results in a separation between the negatively-rotating filament and the other ones. With more filaments,  different numbers of  motors may switch directions, leading to different unbundled states. This is illustrated in Fig.~\ref{UnBu6} in the case of $N=6$ filaments where we show the long-time conformation of the  rotating shapes as a function of the number of negatively-rotating filaments and their location (indicated by a red dot at their base).

\section{Conformation instabilities}
\label{sec:inst}

One of the main results predicted by our model is the occurrence of   instabilities in the  conformation of the filaments. This is best seen by inspecting  Fig.~\ref{hysteresis}. The dimensionless bending energy of the steady-state shape of each filament clearly indicates that sharp transitions occur from weakly bent to crossing ($\Bu\approx 2.1$) and from crossing to bundling ($\Bu\approx 42$). In addition, each transition is associated with a strong hysteresis loop. 

Both  crossing and  bundling instabilities share the same physical origin. Hydrodynamic interactions bend the filaments and are resisted by elastic forces. At a  critical rotation rate, the bending resistance is unable to balance the hydrodynamic stresses, and the filament transition to a new conformation. 

From a kinematic standpoint, if two nearly-straight filaments cross at some point along their length, their crossing is  expected to take place at either a single point or along many points; indeed if two filaments happen to cross at one point, and then are made to cross at a second location along their length, then the remaining portion of the filaments is expected to remain in close continuous contact. It is therefore expected that two crossings implies in fact many crossings.  

\subsection{Crossing instability}

Intuitively, we propose therefore that the fundamental instability to understand  is the first, crossing instability. In this section, by  focusing on the case of two filaments, we present a theoretical approach, together with a simple two-dimensional model, to show how to predict that  instability.

\subsubsection{Small $\Bu$ analysis}

We start by solving the problem for  the steady-state shape of the filaments in the  small-$\Bu$ limit. The steady state of Eq.~\ref{2filaments_components} satisfies the equation
\begin{align}\label{SteadyState_component}
\left[1-\frac{\ln(\epsilon_h d)}{\ln\epsilon_a}\right]	
\frac{\partial^4}{\partial s^4}
\begin{bmatrix}
x\\y	
\end{bmatrix}
=\frac{2\Bu}{d^2}
\begin{bmatrix}
-y\\
x	
\end{bmatrix}.
\end{align}

Based on the symmetries in the geometry, we expect $x$ to be odd in $\Bu$ while $y$ is expected to be even. In the limit of small values of $\Bu$ we thus look to solve Eq.~\ref{SteadyState_component} as a regular  series expansion
\begin{equation}
	x=\Bu x_1+\Bu^3x_3+...,\quad y=-\frac{1}{2}+\Bu^2y_2+...,
\end{equation}
and we aim to solve for the leading order deflections $(x_1,y_2)$.  
In order to proceed with the solution, we also need to choose values for the small dimensionless parameters $\epsilon_a$ and $\epsilon_h$. Given the numbers relevant to the bundling of bacterial flagella discussed in  Sec.~\ref{sec:setup}, we choose the relevant values  $\epsilon_a=0.001$ and $\epsilon_h=0.1$.

With these assumptions, the solution of Eq.~\ref{SteadyState_component} at  $O(\Bu)$ is 
\begin{equation}\label{x4stip}
\frac{\partial^4x_1}{\partial s^4}=\frac{3}{2}\cdot
\end{equation}
With  the boundary conditions given  in Eq.~\ref{BCs}, we obtain the solution as
\begin{equation}
x_1=\frac{s^4}{16}-\frac{s^3}{4}+\frac{3s^2}{8}\cdot
\end{equation}
At next order,  $O(\Bu^2)$, we have to solve
\begin{equation}
\frac{\partial^4y_2}{\partial s^4}=3x_1,	
\end{equation}
whose solution satisfying the boundary conditions in Eq.~\ref{BCs} is
\begin{equation}\label{ytip}
y_2=	\frac{s^8}{8960}-\frac{s^7}{1120}+\frac{s^6}{320}-\frac{3s^3}{80}+\frac{13s^2}{160}\cdot
\end{equation}

\begin{figure}[t]
\centering
\includegraphics[width=0.85\textwidth]{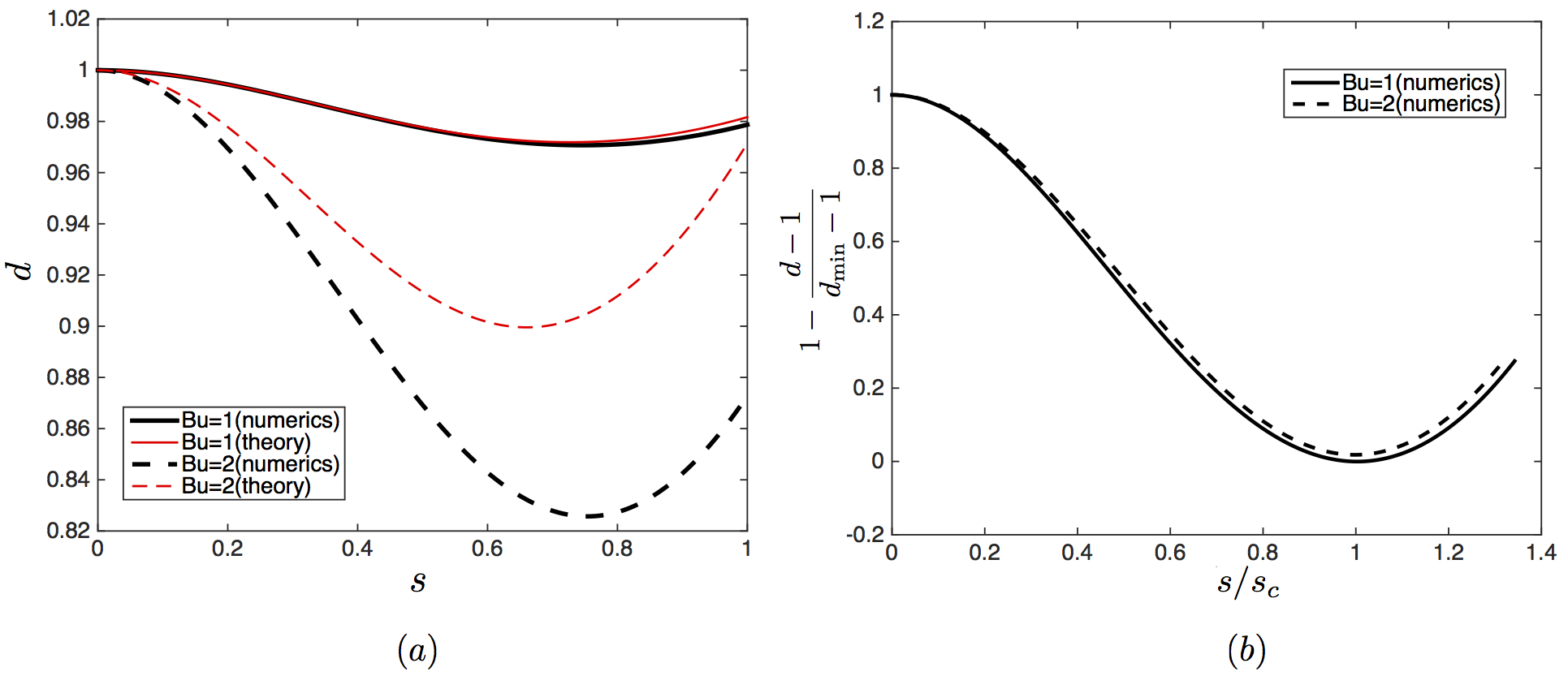}
\caption{($a$) Comparison between  the asymptotic solution for the deflection of the filaments  (thin red lines) with the  numerical solution (thick black lines) for $\Bu=1$ (solid) and $\Bu=2$ (dashed); 
 ($b$) A closer look at the numerical solutions for for $\Bu=1$ (solid) and $\Bu=2$ (dashed) indicated that up to a re-scaling they have the same shape ($s_c$ is the  value of the arclength at which the distance between the filaments is  minimum).}\label{base}
\end{figure}

We compare the numerical solution  to the  asymptotic solution in Fig.~\ref{base}a. We find that for $\Bu=1$, the agreement with between the two is very good; however for $\Bu=2$ (close to the instability point), the  difference between the computational solution and the theoretical one is more important. 

\subsubsection{Analytical ansatz}

In order to build  a more accurate analytical model for the deflection of the filaments, we need to correct the asymptotic solution so it remains valid up to the instability point. To do so, we examine the shape of the filaments as obtained numerically for $\Bu=1$ and $\Bu=2$ and plot the rescaled distance between the filaments  in Fig.~\ref{base}b as a function of the arclength scaled by the arclength where the minimum distance occurs, $s_c$, by linearly mapping $d$ 
so it  reaches zero  at the minimum point  i.e.~to
\begin{equation}
1-\displaystyle\frac{d-1}{d_{\rm min}-1}\cdot
\end{equation}
 We see that, up to a rescaling, the shapes at $\Bu=1$ and $\Bu=2$ are identical and the only difference is how their magnitude scale with the change in $\Bu$. 

\begin{figure}[t]
\centering
\includegraphics[width=0.85\textwidth]{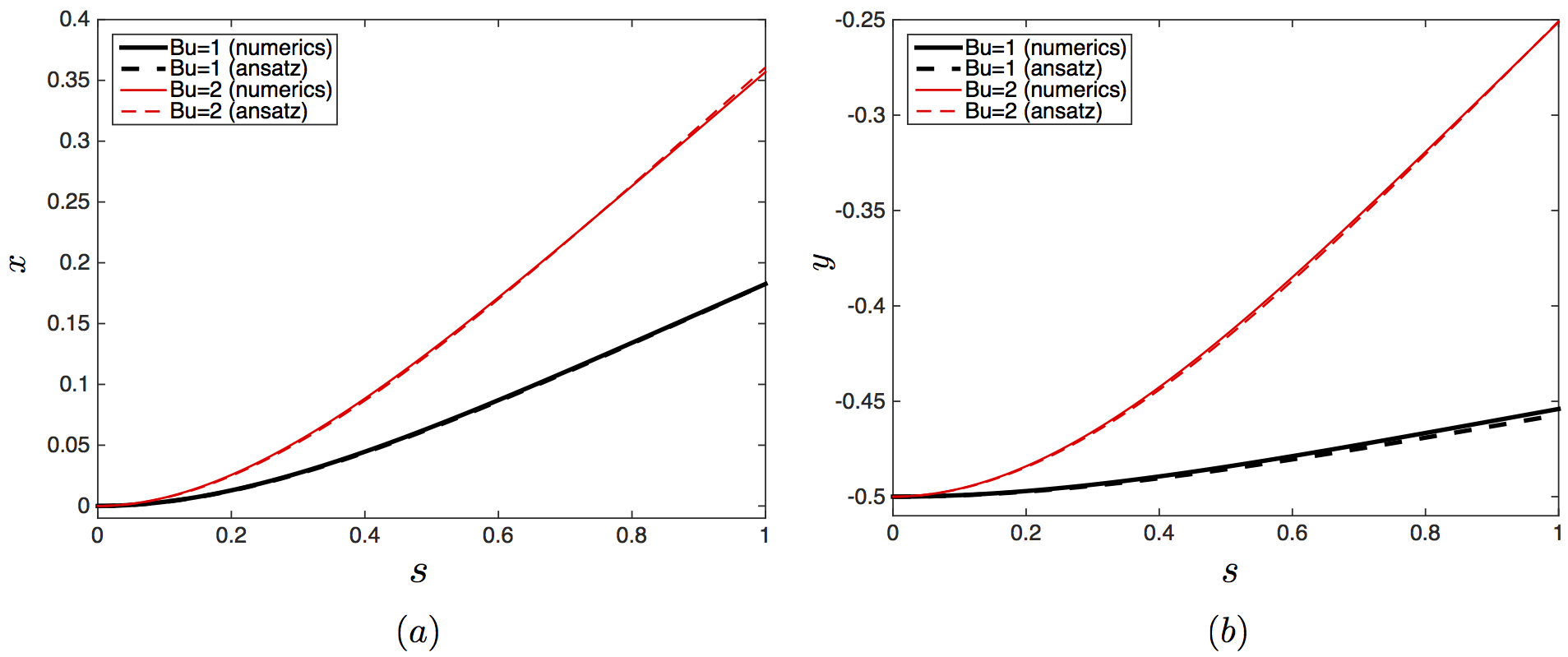}
\caption{Comparison between the numerical steady shapes (solid lines) and the shape predicted by the analytical ansatz (dashed line):  
$(a)$: $x(s)$;  $(b)$:  $y(s)$. The shapes are compared for $\Bu=1$ (thick black lines) and $\Bu=2$ (thin red lines).}\label{comparetoNum}
\end{figure}

To construct a better analytical model, we then proceed by choosing a steady reference state $[\Bu_o, x_o(s), y_o(s)]$ where the asymptotic solution  provides a good estimate  (we choose $\Bu_o=1$). We denote by $[s_{oc}, x_{oc}, y_{oc}]$  the critical point where the distance between the filaments, $d$, reaches its  minimum value. We then numerically track this same critical point as we increase the value of $\Bu$, which we denote $[s_c(\Bu), x_{c}(\Bu), y_{c}(\Bu)]$. From this, and given the similarity of the shapes seen in Fig.~\ref{base}b, we can construct an analytical steady-state ansatz $[\Bu, x(\Bu; s), y(\Bu; s)]$ by simply scaling in $x$, $y$ and $s$   the reference analytical solution as
\begin{subeqnarray}\label{guess}
x(\Bu; s)&=&\frac{x_c}{x_{oc}}x_o\left({\rm Bu}_o; \frac{s_{oc}}{s_{c}}s\right)\\
y(\Bu; s)&=&-\frac{1}{2}+\frac{y_c+\frac{1}{2}}{y_{oc}+\frac{1}{2}}\left[y_o\left({\rm Bu}_o; \frac{s_{oc}}{s_c}s\right)+1/2\right].
\end{subeqnarray}

We show in Fig.~\ref{comparetoNum} a comparison for each component of the shapes 
$x(s)$  (Fig.~\ref{comparetoNum}a) and $y(s)$  (Fig.~\ref{comparetoNum}b) between the computations (solid lines) and the analytical ansatz (dashed lines). We see that the ansatz, obtained analytically for $\Bu=1$, is able to fully capture the shape obtained numerically at $\Bu=2$. 

\subsubsection{Linear stability of analytical ansatz}

We may then use  this analytical ansatz as an accurate base state, $[x_b(\Bu; s), y_b(\Bu; s)]$,   around which we may  carry out a linear stability calculation. Assuming small deviations around the base state and exponential growth, we decompose the general  shape  $[x(s,t), y(s,t)]$ of the filament  as
\begin{equation}
x(s,t)=x_b(\Bu;s)+	\hx (s)e^{\sigma t},\quad y(s,t)=y_b(\Bu;s)+\hy(s) e^{\sigma t},
\end{equation}
and substitute into Eq.~\ref{SteadyState_component}, leading to the linear system,
\begin{align}\label{eq:linear}
\begin{split}
\sigma
\begin{bmatrix}
\hx\\ \hy
\end{bmatrix}&+
\begin{bmatrix}
\displaystyle\frac{2}{3}-\displaystyle\frac{\ln 2\sqrt{x_b^2+y_b^2}}{\ln\epsilon_a}
\end{bmatrix}
\begin{bmatrix}
\hx\\ \hy
\end{bmatrix}_{4s}
=\mathbf{A}
\begin{bmatrix}
\hx\\ \hy
\end{bmatrix},
\end{split}
\end{align}
where the matrix $\bf A$ is given by
\begin{align}
\begin{split}
\mathbf{A}=
\begin{bmatrix}
\displaystyle\frac{x_b(x_b)_{4s}}{(\ln\epsilon_a)(x_b^2+y_b^2)}+\frac{\Bu x_by_b}{(x_b^2+y_b^2)^2}&\quad& \displaystyle\frac{y_b(x_b)_{4s}}{(\ln\epsilon_a)(x_b^2+y_b^2)}
+\frac{\Bu(y_b^2-x_b^2)}{2(x_b^2+y_b^2)^2}\\\\
\displaystyle\frac{x_b(y_b)_{4s}}{(\ln\epsilon_a)(x_b^2+y_b^2)}
+\frac{\Bu(y_b^2-x_b^2)}{2(x_b^2+y_b^2)^2} &\quad& \displaystyle\frac{y_b(y_b)_{4s}}{(\ln\epsilon_a)(x_b^2+y_b^2)}-\frac{\Bu x_by_b}{(x_b^2+y_b^2)^2}
\end{bmatrix}.
\end{split}
\end{align}

\begin{figure}[t]
\centering
\includegraphics[width=3.6in]{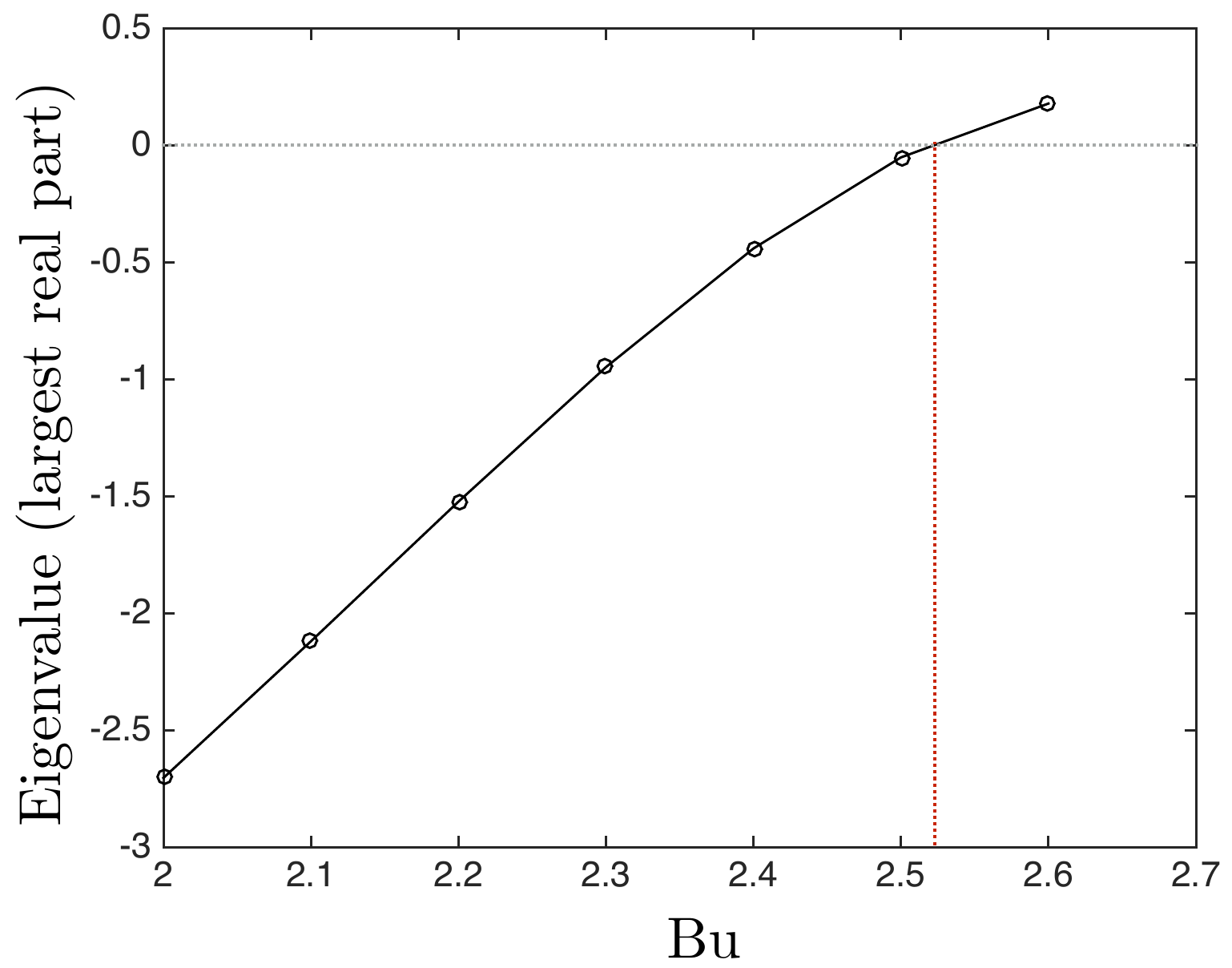}
\caption{Solution of the eigenvalue problem in Eq.~\ref{eq:linear} with the largest real part as a function of the Bundling number, $\Bu$. The red dotted dash line indicated the critical point of transition from stable to unstable, which takes places slightly above $\Bu=2.5$.}\label{EIVs}
\end{figure}

We  use fourth-order finite differences    in order to solve numerically the  eigenvalue problem in Eq.~\ref{eq:linear}. We plot in Fig.~\ref{EIVs} the eigenvalue with the larges real part as a function of the Bundling number. The system is seen to become linearly unstable slightly above $\Bu=2.5$, a  value    close to the computational result of $\Bu\approx 2.1$ for the crossing instability.

\subsubsection{Two-dimensional toy model}

The analytical asymptotic solution with  rescaling of its amplitude allows to capture the shape instability at a critical value of the Bundling number, and thus the transition to the crossing configuration. In order to understand intuitively the physics behind this instability, we consider a simple toy problem displaying the same instability. 

\begin{figure}[t]
\centering
\includegraphics[width=6in]{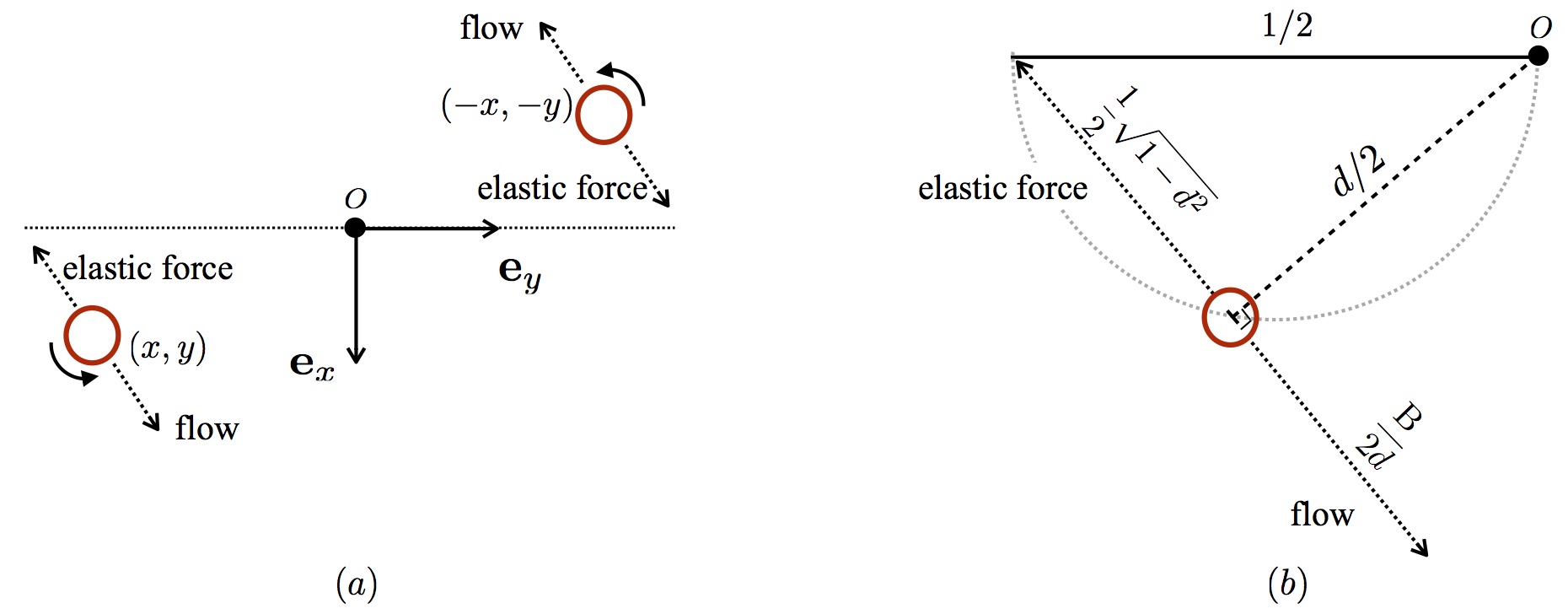}
\caption{
($a$) Setup for two-dimensional cylindrical model: two parallel rigid  cylinders  linked elastically to a reference position interact hydrodynamically;  
($b$) Geometric calculation of magnitude of forces (see text for details).}\label{BeadsModel}
\end{figure}

Instead of a continuous filament, consider two parallel rigid  cylinders which are linked elastically to a reference position. The cylinders are assumed to rotate at a constant rate in the fluid and thus to  interact hydrodynamically. This setup is illustrated schematically in Fig.~\ref{BeadsModel} with the same axis notation as for the two-filament case.

We denote the location of  cylinder \#1 by $(x, y)$ in Cartesian coordinates. At $t=0$, it is assumed to be at  $(x, y)=(0, -1/2)$. By symmetry,  cylinder \#2 is located at $(-x, -y)$. Assuming a elastic force simply proportional to the displacement of the cylinders away from their reference configuration, its magnitude is simply given by $(x, y+1/2)$ for cylinder \#1.
The force balance between local viscous drag, elastic restoring force and hydrodynamic interactions leads therefore to the dimensionless equation for the position $(x,y)$ of cylinder \#1 as
\begin{align}
\frac{\partial}{\partial t}
\begin{bmatrix}
	x\\ y
\end{bmatrix}
	+
    \begin{bmatrix}
    x\\ y+\frac{1}{2}	
    \end{bmatrix}
    =\frac{\rm B}{d^2}
    \begin{bmatrix}
    	-y\\
    	x
    \end{bmatrix},
\end{align}
where $\rm B$ is a dimensionless coefficient similar to the Bundling number  which includes  geometry and elasticity, and $d=2\sqrt{x^2+y^2}$ is the distance between cylinders.

For which values of $\rm B$ can a steady state be found? At steady state, the hydrodynamic force has to balance the elastic resistance. If both cylinders are located at a distance $d$ from each other, the magnitude of the hydrodynamic force is given by ${\rm B }/2d$ while the elastic restoring force is given by the distance to the reference point. Since the hydrodynamic force acts at right angle to the vector joining the cylinders, each cylinder is constrained mechanically to remain  on a circle of diameter $1/2$ (see Fig.~\ref{BeadsModel}b). the distance between the cylinder and its reference configuration 
 may then be obtained using the Pythagorean Theorem as 
$\sqrt{{1}/{4}-{d^2}/{4}} = \sqrt{1-d^2}/2$. Force balance between elasticity and hydrodynamics requires therefore that 
\begin{equation}
{\rm B}=d\sqrt{1-d^2}.
\end{equation}
The right hand-side of this equation is bounded from above by $1/2$ and thus there exists a steady state only for ${\rm B} \leq {1}/{2}$, which corresponds to a finite critical distance between the cylinders of $d=1/\sqrt{2}$.  This simple elasto-hydrodynamic model  allows therefore to reproduce the same physics of a continuous elastic deformation induced by hydrodynamic interactions until a critical finite distance and a shape bifurcation.

\subsubsection{Continuous anaytical model}
Inspired by the success of the  simple cylindrical model, we adapt a similar idea to the case of two elastic filaments by focusing on the elastic deformation at their tips. Using the asymptotic results in Eqs.~\ref{x4stip} - \ref{ytip}, we see that for small values of $\Bu$  the components at tip ($s=1$) of the elastic forces and the displacements are given by
 \begin{align}
&x_{4s}(s=1)=\frac{3}{2}\Bu,\quad x(s=1)=\frac{3}{16}\Bu,\\
 &y_{4s}(s=1)=\frac{9}{16}\Bu^2,\quad y=-\frac{1}{2}+\frac{413}{8960}\Bu^2.	
\end{align}
Using these results, we can then infer a linear force-displacement relationship valid at the tip at small $\Bu$
\begin{equation}\label{linear}
x_{4s}=8x,\quad y_{4s}=\frac{5040}{413}\left(y+\frac{1}{2}\right).
\end{equation}
Using Eq.~\ref{linear} to replace the continuous elastic forcing by one proportional to the displacement of the tip away from its reference point, we may replace Eq.~\ref{SteadyState_component} by an algebraic equation as
\begin{align}\label{vectors}
\left[1-\frac{\ln(\epsilon_h d)}{\ln\epsilon_a}\right]	
\begin{bmatrix}
8x\\ \displaystyle\frac{5040}{413}\left(y+\frac{1}{2}\right)
\end{bmatrix}
=\frac{2\Bu}{d^2}
\begin{bmatrix}
-y\\
x	
\end{bmatrix}.
\end{align}
Since the two vectors in Eq.~\ref{vectors} are proportional to each other, their cross product is zero leading to the identity
\begin{equation}\label{x}
x^2=-\frac{630}{413}y\left(y+\frac{1}{2}\right)\cdot
\end{equation}
As a consequence we have 
\begin{equation}\label{d}
\frac{d^2}{4}=x^2+y^2=-\frac{217}{413}y^2-\frac{315}{413}y.	
\end{equation}
We then use the first row of Eq.~\ref{vectors} to obtain $\Bu$ as an explicit function of $y$
\begin{equation}\label{Bulimit}
\Bu=-\left(1-\frac{\ln \epsilon_h d}{\ln\epsilon_a}\right)\frac{4xd^2}{y}	,
\end{equation}
with $x$ given by Eq.~\ref{x} and $d$ by Eq.~\ref{d}.  

\begin{figure}[t]
\centering
\includegraphics[width=3.5in]{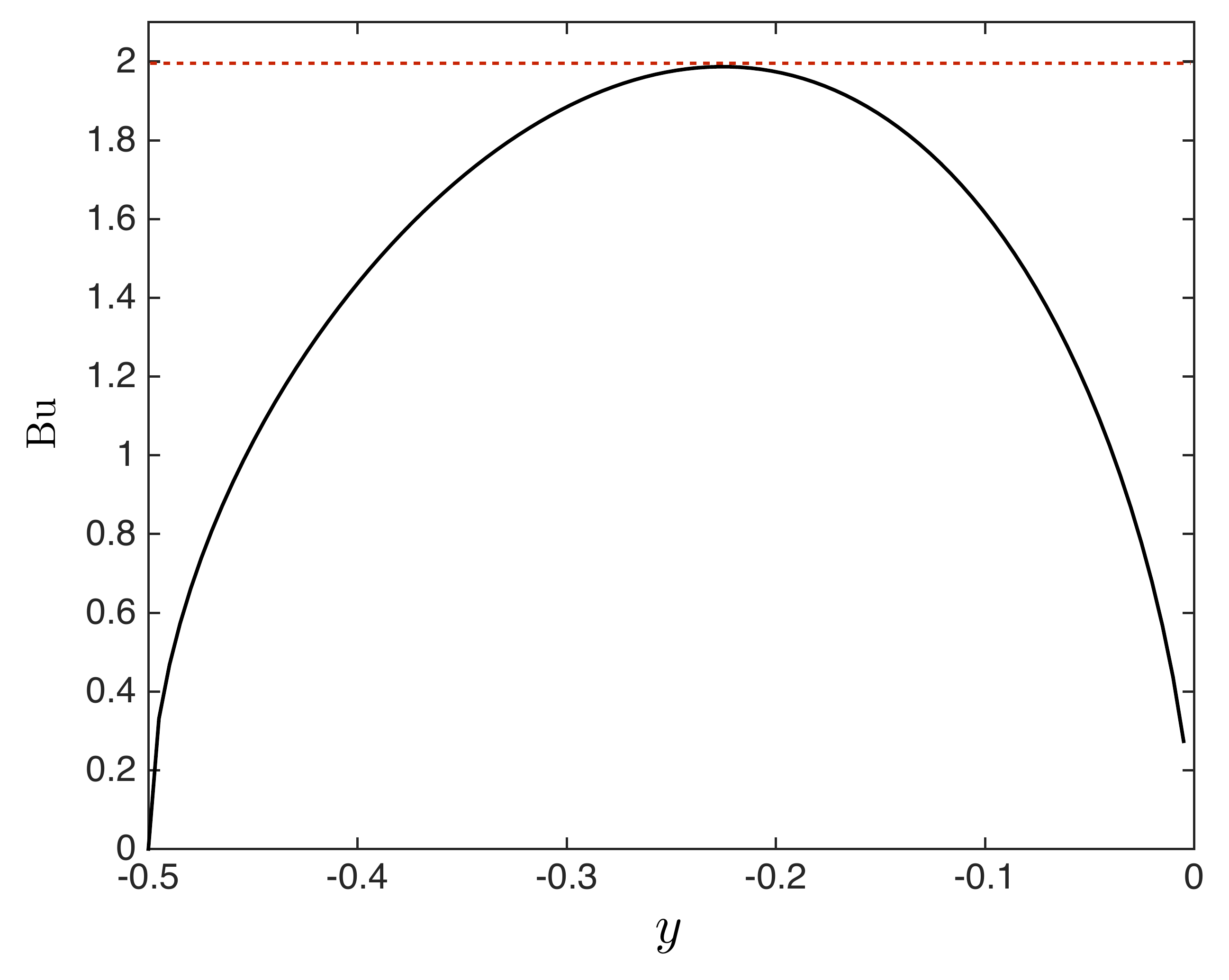}
\caption{Numerical value of  $\Bu$ as a function of $y$ according to the analytical model,  Eq.~\ref{Bulimit}, showing a maximum possible Bundling number of approximately 2.
}\label{SingleModel}
\end{figure}

The value of the right-hand side of this equation may be computed for all $y$ in the interval $-0.5 \leq y<  0$, with results shown in Fig.~\ref{SingleModel}. 
We see that $\Bu$ reaches its maximum value around $\Bu 
\approx 2$ at position $(x,y)=(0.31, -0.23)$, corresponding to the tip-to-tip  distance $d\approx 0.76$ (note that this maximisation may also be done by hand, leading to a transcendental equation due to the presence of logarithms).   When $\Bu$  is larger than this critical value, no steady-state solution exists,  and the confirmations of the filaments transition to a new state. This analytical estimate ($\Bu 
\approx 2$)  agrees very well with the critical value of $\Bu$ found numerically ($\Bu 
\approx 2.1$) and allows therefore to successfully capture the physics of the crossing instability.

\subsection{Bundling instability}
In order to propose an estimate for the critical Bundling number at the second instability (from crossing to bundled state), we need to examine the dependence of $\Bu$ on the physical parameters of the problem. We recall that $\Bu$ is defined by
\begin{equation}\label{117}
	{\rm Bu}= \frac{\xi_\perp\omega_0  a^2 L^4}{h_0^2A}\cdot
\end{equation}
The important point from Eq.~\ref{117} is that if the distance between the filaments, $h$, is made to increase or if the length of the filaments, $L$, is made to decrease, the critical rotation rate needed to get an instability has to increase in order to compensate for it. 

Examining the typical shape of the filaments in the crossing configuration (see Fig.~\ref{SteadyStates}), we see that the crossing point it located about one third of the way along the filament, so that the end-to-end distance between the free ends of the filaments is about twice the original distance at their clamped ends. If it is true that the physics responsible for the crossing $\leftrightarrow$  bundling transition is similar to the one responsible for the bent $\leftrightarrow$ crossing, we can then propose a  heuristic  estimate for the second critical $\Bu$ number. We approximate the crossing configuration as a new initial configuration with $L$ decreased to $2L/3$ and $h_0$ increased to $2h_0$, i.e.~we consider only the shape after the crossing point. That shape  is expected to display a conformation instability when its  Bundling number us  approximately equal to 2.1 (i.e.~the value for the first instability), meaning that we expect
\begin{equation}
\frac{\xi_\perp\omega_0  a^2 (2L/3)^4}{(2h_0)^2A}
\approx 2.1 ,
\end{equation}
which leads to the estimate for the original $\Bu$ number at the second transition as
\begin{equation}
\Bu=\frac{\xi_\perp\omega_0  a^2 L^4}{h_0^2A}\approx 42.5.
\end{equation}
This estimate is in very good agreement with our numerical results,  conforming {\it a posteriori} the similarity of the physics for the two transitions. Note that the configuration of the filaments past the crossing state does not satisfy the separation of scales assumed in our original theory, so the agreement between the estimate above and our numerical simulations need to be taken as  qualitative at best.

\section{Experiments}

\begin{figure}
\centering
\includegraphics[width=0.8\textwidth]{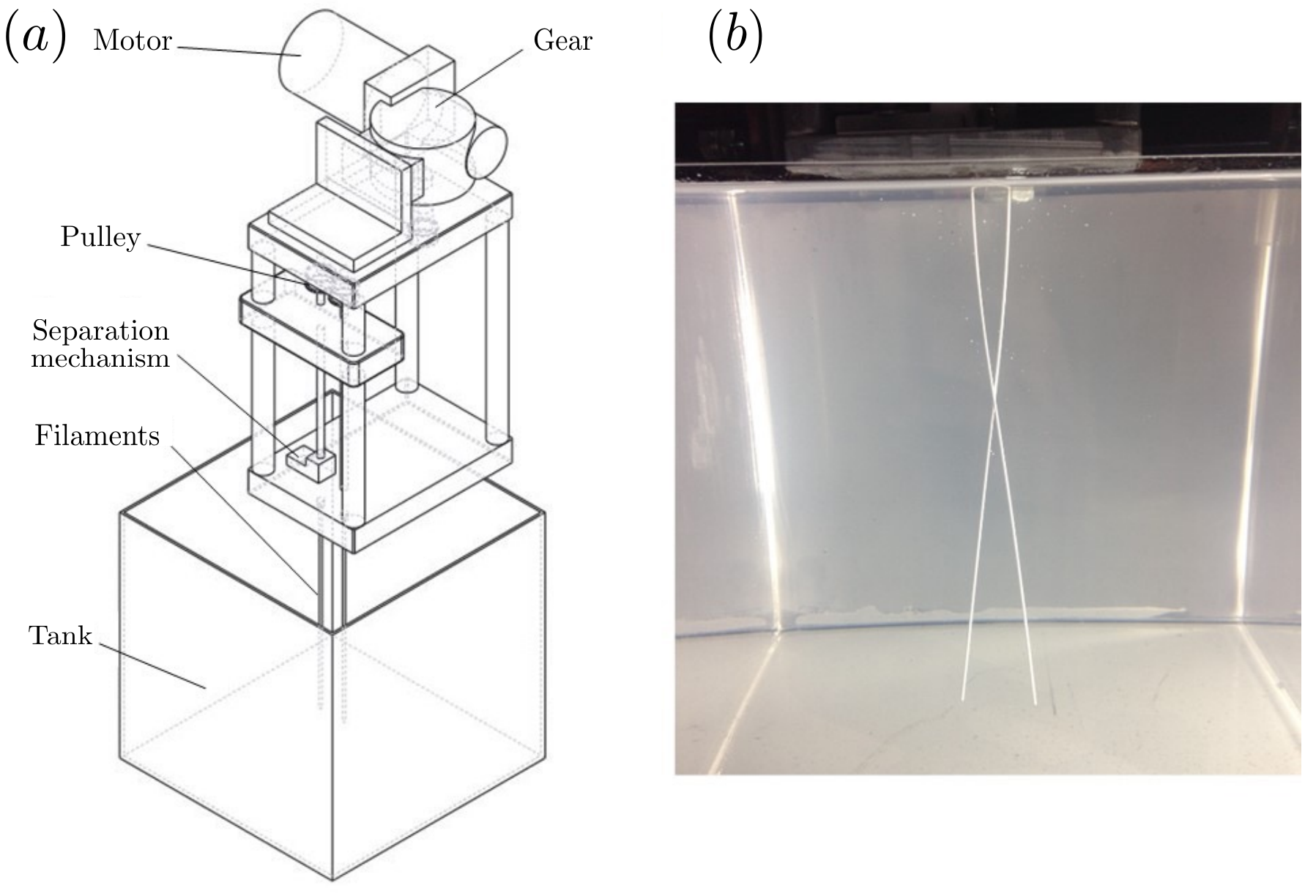}
\caption{(a) Schematic of the experimental arrangement; (b) Illustrative image of crossing state (note that the driving in the experiments is located at the top, and thus the experimental pictures are inverted vertically compared to the numerical shapes shown earlier).}
\label{fig18}
\end{figure}

\subsection{Setup}

In order to validate the analytical and numerical results described herein, a bespoke experimental facility was designed and built, a schematic of which is shown in Fig.~\ref{fig18}a.  The rig utilises a geared motor to drive a shaft which, in turn, uses a belt and pulley system to rotate two interchangeable stainless steel rods, of diameter 1.6~mm, each housed in its own bearing in the top supporting structure.  The motor is connected to a digital encoder which provides a direct read out of the rotation speed $\omega_0$.  The separation of the two stainless steel rods can be adjusted at the bottom supporting structure from a minimum of 10~mm to a maximum of 20~mm in steps of 1~mm.  To ensure the accuracy of this separation, spacers of known separation were machined and used to set the filament spacing.  At the downstream end of the stainless steel rods, a 1~mm hole was drilled into each into which elastic filaments of various radii and length could be securely bonded.  The elastic filaments were made of polystyrene (``Evergreen StripStryene'' supplied by wonderland models) and had nominal diameters ($2a$) of 0.02, 0.025, 0.03, 0.04 inches (508, 635, 762 and 1016~$\mu$m respectively) and original length 350~mm.  The Young's modulus of the filaments $E$ was measured using an Instron tensile test machine and found to be 1~GPa ($\pm$0.02~GPa) up to 0.5\% strain.  The diameter of each individual filament was measured using a digital micrometer at three locations along its length and the average radii was found to be consistent within $\pm$10~$\mu$m or better.  The filaments were cut to specific lengths in the range $140-180$~mm with an accuracy of $\pm$0.5~mm.  These filaments where rotated in a cubic bath ($20\times20\times20$~cm$^3$ in volume) of silicon oil (of viscosity 12.8~Pa.s at 20$^\circ$C supplied by Basildon Chemical Co. Ltd) and the temperature of each experiment measured using a mercury thermometer.  An Anton Paar MCR 302 controlled-stress rheometer was used to directly measure the viscosity of the silicon oil over the range of different temperatures encountered.  During a typical run the temperature of the fluid would change by less than 0.25$^\circ$C and all experiments were conducted between 19 and 21.5$^\circ$C.  

\begin{figure}
\centering
\includegraphics[width=0.7\textwidth]{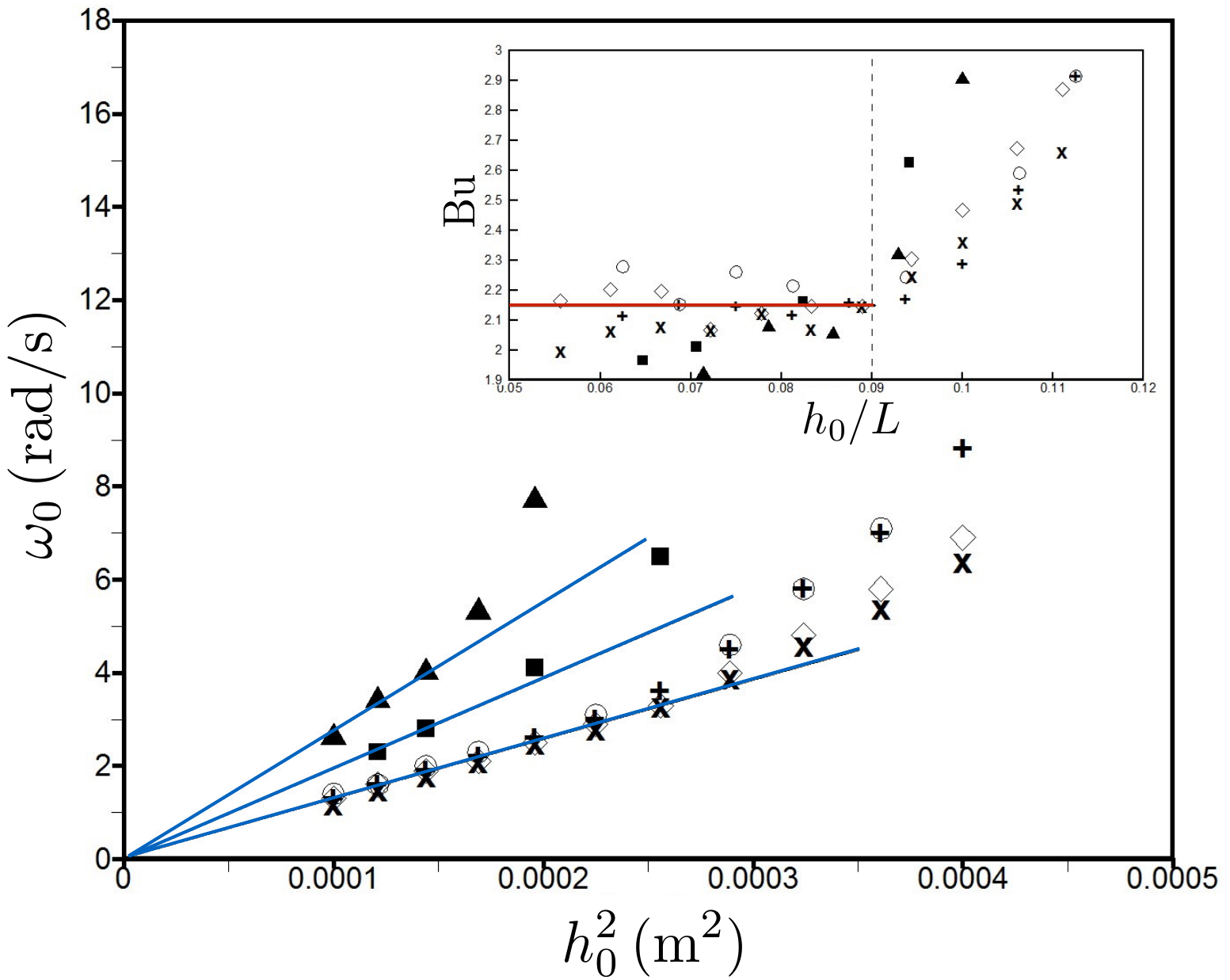}
\caption{Critical angular velocity, $\omega_0$, versus $h_0^2$ for first crossing instability. 
({$\lozenge$,$\times$}): filament length $L =180$~mm, filament radius $a =0.44$~mm. 
({$\blacktriangle$}):     $L =140$~mm,   $a =0.41$~mm. 
({+,{\Large$\circ$}}):   $L =160$~mm,   $a =0.35$~mm. 
({$\blacksquare$}):   $L =170$~mm,   $a =0.49$~mm.  
For all data, the elastic modulus of the filaments is $E = 1$~GPa and the viscosity of silicon oil is 12.8 Pa.s (at 20$^\circ$C). Straight lines are linear best fits to the data for $h_0/L < 0.09$. Inset shows same data converted to critical Bundling number variation versus $h_0/L$. For $h_0/L< 0.09$, the results are consistent with a critical value of $\Bu = 2.15 \pm0.2$.}
\label{fig19}
\end{figure}

\subsection{Experimental results}

A  number of different experiments were conducted with all results  shown in Figs.~\ref{fig19}.  A filament of given radius ($0.35 -0.44$~mm) was selected,  cut to a given length (140-180 mm) and then placed into the silicon oil at the smallest separation possible (10 mm).  These values were selected to try to ensure that $a \ll h_0 \ll L$ (actual range $0.0175 < a/h_0 < 0.044$ and  $0.056 < h_0/L < 0.143$).  Care was taken to guarantee that the filaments were initially parallel within the fluid.    The filament would then be rotated at a low angular velocity (less than 0.5 rad/s) and allowed to reach a steady-state configuration over a period of about 15 minutes.  If the filaments remained parallel the angular velocity would then be incremented by a small amount (about 0.5 rad/s) and the process repeated until the crossing state -- illustrated in Fig.~\ref{fig18}b -- was observed.  At this point the motor would be switched off, the filaments physically separated and realigned.  The experiment would then be restarted at an angular velocity slightly below the critical value and allowed to rotate for an extended period of time to confirm the critical rotation speed.  Once the critical value of $\omega_0$ was confirmed, the filament separation would be incremented by 1~mm and the process repeated.  In this manner, data sets of critical angular velocity for different filament radii, length and separation could be obtained.

In Fig.~\ref{fig19}, this data is plotted as critical angular velocity, $\omega_0$,  versus $h_0^2$ as the analytical and numerical results predict that this should be a linear relationship all else being held equal (see Eq.~\ref{Bu-dimensional}).  As can be seen in Fig.~\ref{fig19}, for sufficient small values of $h_0/L$, different data sets exhibit a good linear scaling (Pearson correlation R varying from 0.985 to 0.9985).  When converted to a critical Bundling number, as shown in the inset to Fig.~\ref{fig19}, this is seen to be approximately constant for all data sets at $\Bu=2.15\pm 0.09$  in excellent agreement with the theoretical and numerical values, provided that $h_0/L$ remains smaller than approximately 0.09 (recall that the theory was derived in the asymptotic limit $h_0/L\ll 1$).  The small degree of scatter in the value of critical Bundling number arises from low values of critical angular velocity which, at these values, fluctuates somewhat ($\pm$10\%) as the motor is employed at the lowest end of its working range.  Nevertheless, the repeatability of the results is good as is illustrated in Fig.~\ref{fig19}  for two nominally identical experiments and the linear scaling appears robust across different filaments.

\begin{figure}
\centering
\includegraphics[width=0.5\textwidth]{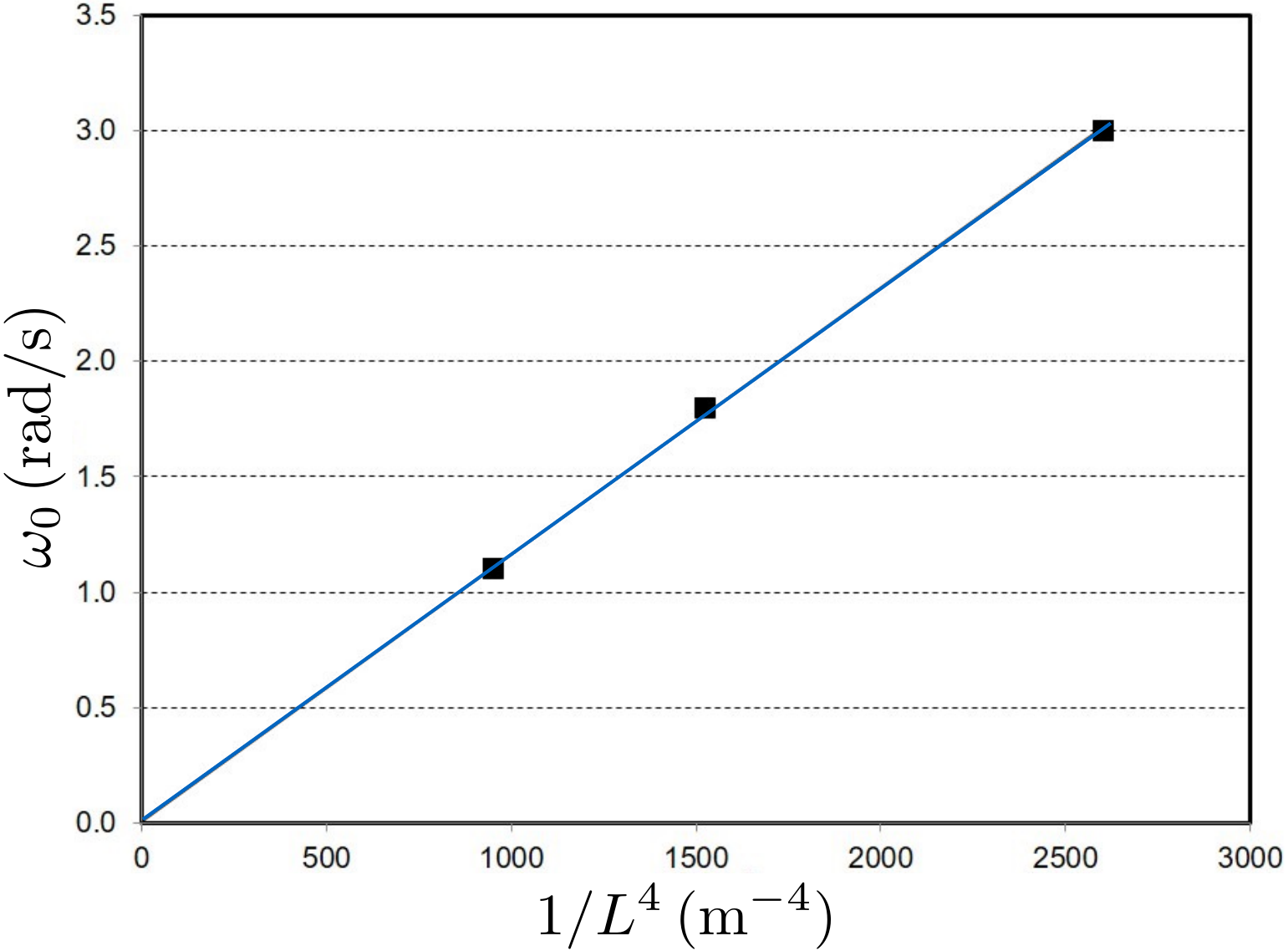}
\caption{Plot of critical angular velocity, $\omega_0$, versus $1/L^4$ for first crossing instability. For all data, filament radius a = 0.42 mm, filament separation $h_0 = 10$~mm, elastic modulus of filaments $G = 1$~GPa and viscosity of silicon oil of 12.8~Pa.s (at 20$^\circ$C). Straight line is linear best fit ($R=0.9993$) to the data giving a constant critical $\Bu=2.14$.}
\label{fig20}
\end{figure}

In Fig.~\ref{fig20} we  show how the critical angular velocity varies with $1/L^4$ for the smallest separation possible ($h_0=10$~mm) to ensure that $h_0/L$ remained in the asymptotic regime ($0.056 < h_0/L < 0.071$).  Once again an excellent linear scaling ($R =0.9995$) is observed in agreement with the theory.  For this experiment the same filament was utilised, being successively cut to shorter lengths, to ensure that the filament radius remained constant.  The critical Bundling number in this case is $\Bu=2.14$ being, once again, in excellent agreement with theoretical and numerical results. 

We note that the  rig was designed such that both the crossing and bundling instabilities could be studied, however the larger filament deformations observed in the bundling instability resulted in filament failure/breakage at the point where it was bonded into the stainless steel rod.  Such failure is illustrated in the supplementary material where a video of the phenomena shows how the filaments react to a constantly increasing angular velocity \cite{supp_rob}.  As a consequence, we restricted our interest here to the first crossing instability.  We confirmed that the maximum Reynolds number ($=\rho a^2\omega_0/\mu $ where $\rho$ is the density of the fluid) remains vanishingly small being, at most, about $2\times10^{-4}$ for the largest diameter filament at the highest angular velocity studied.  Finally, the possibility of hysteric behaviour was briefly examined and a marked effect was observed such that the filaments remained in the crossed state at lower angular velocities when approaching ``from above'', i.e. slowly reducing the angular velocity from an already crossed state.  Both effects form part of a larger experimental study currently underway and will be reported on in detail elsewhere.

\section{Conclusion}

In this paper, inspired by  the bundling of bacterial flagella, we proposed  a simple model of flow-induced wrapping of flexible filaments. By deriving at leading-order the  hydrodynamic interactions between two nearby straight elastic filaments of radius $a$ and length $L$ and separation $h_0$  satisfying $a\ll h_0\ll L$, we have obtained a partial differential equation describing the shape of the filament  controlled by a single  dimensionless  Bundling number, $\Bu$. Based on this model, we  studied the bundling and unbundling dynamics of two or more filaments. We showed that the filaments undergo two conformation instabilities from weakly bent to crossing and then to bundled at two critical values of $\Bu$.    
 We analytically tackled the instability occurring in the first transition by building a small $\Bu$ ansatz and a simplified two-dimensional model, leading respectively to predicted critical values of  $\Bu=2.5$ and 2, in good agreement with the numerically computed value of $\Bu=2.1$.  We then used a very simple estimate to predict the occurrence of second instability obtained numerically at a much large value of $\Bu\approx 42$.   In order to validate our model and theoretical approach, we then carried out macro-scale experiments using polystyrene  rods rotated in silicon oil. The experimental results were in excellent agreement with the scalings predicted by the  theory, and in particular we measured a critical value of  $\Bu=2.15 \pm 0.2$ for the transition to the crossing instability.

Beyond the far-field limit in which our  model was derived, $h_0\gg a$,  the configuration where the  filaments almost touch each other has yet to be fully modelled.  In our computations, we have used repulsive interactions in order to  keep the minimum distance between the filaments on the same order of magnitude as the radius $a$ (never less than three radii in practice).  In the lubrication limit of nearly-touching filaments,  hydrodynamic forces scale as $O(a^{3/2}h_0^{-1/2})$, while in our calculation the far-field result offers scaling of $O(a^2/h_0)$. In the cases where $h_0\sim a$ as in our work, both forces are $O(a)$. If the filaments were made to approach any further, the far-field hydrodynamic forces should be replaced by much smaller lubrication forces, and a more sophisticated model would be required to untangle the matching of lubrication with far-field solutions.

The framework  developed in this paper   
 to calculate  hydrodynamic interactions between filaments  is very general and could be extended to a number of   situations where  nearby slender filaments interact through viscous fluids.  The next logical step in the theoretical investigation of bacterial bundling would be to quantify the role played by the helical geometry of bacterial flagellar filaments.   
In that case, a  third length scale enters the analysis, namely the radius of of the helix. As a result, and notably different from the  straight case address here, an axial force  will be generated by each rotating helix, which will impact the flow generated and the dynamics of bundling.  Beyond the application to bacteria, we expect that our framework will be useful to tackle a wide range of problems in soft matter and biological physics such as the flow of flexible filaments, the dynamics of cilia arrays and cytoskeletal mechanics.

\section*{Acknowledgements}
This work was supported in part by the  Cambridge Overseas Trust (YM), a  Marie Curie CIG grant (EL) and an ERC consolidator grant (EL).

\bibliographystyle{unsrt}

\end{document}